\documentclass[aps,prd,preprint,groupedaddress,showpacs]{revtex4}

\usepackage{epsfig}
\usepackage{graphics}
\usepackage{slashed}
\usepackage{color}
\usepackage[citecolor=true]{hyperref}
\usepackage{multirow}
\usepackage{amsmath}
\begin{document}

\leftmargin -2cm
\def\choosen{\atopwithdelims..}

\boldmath
\title{Estimates for the single-spin asymmetries  in $p^{\uparrow}p \to J/\psi X$ process\\
at PHENIX RHIC and SPD NICA} \unboldmath

\author{\firstname{A.V.}\surname{Karpishkov}} \email{karpishkoff@gmail.com}
\author{\firstname{V.A.}\surname{Saleev}} \email{saleev@samsu.ru}

\affiliation{Samara National Research University, Moskovskoe Shosse,
34, 443086, Samara, Russia}

\affiliation{Joint Institute for Nuclear Research, Dubna, 141980
Russia}

\author{\firstname{M.A.} \surname{Nefedov}} \email{nefedovma@gmail.com}
\affiliation{Samara National Research University, Moskovskoe Shosse,
34, 443086, Samara, Russia}

\begin{abstract}
We study the transverse single-spin asymmetry (TSSA) in $p^{\uparrow}p
\to J/\psi X$ reaction, incorporating both transverse-momentum and spin effects.
To predict production cross section of prompt $J/\psi$ we use two
different approaches, the  non-relativistic QCD (NRQCD) factorization approach and
the Improved Color Evaporation Model (ICEM), and show how the
predicted results for TSSAs depend on choice of hadronization model.
For initial-state factorization we consider two models: the standard Generalized Parton Model (GPM) and the Colour Gauge-Invariant version of it (CGI-GPM).
Estimates for the TSSAs in $p^{\uparrow}p \to
J/\psi X$ process for the conditions of the future SPD NICA experiment are presented for the first time.
\end{abstract}



\maketitle

\section{Introduction}
\label{sec:Int} The transverse-momentum dependent (TMD) parton distribution
functions (PDFs)  incorporate information about
three-dimensional structure of proton and it's spin properties
\cite{Collins:1989gx,Angeles-Martinez:2015sea}. Among the
leading-twist TMD PDFs, the Sivers function
\cite{Sivers:1989cc,Boer:1997bw} is one of the most interesting and it is widely
investigated in $p^{\uparrow}p \to h X$ inclusive processes
\cite{Anselmino:2013rya,DAlesio:2015fwo}. Sivers function describes the number
density of unpolarized gluons $g$ (or quarks $q$) with intrinsic
transverse-momentum ${\bf q}_T$ inside a transversely polarized
proton $p^{\uparrow}$ , with three-momentum $\bf P$ and spin
polarization vector $\bf S$,
\begin{equation}
F^{\uparrow}_g(x,{\bf q}_T)=F_g(x,q_T)+\frac{1}{2}\Delta^N
F_g^{\uparrow}(x,q_T){\bf S}\cdot (\hat{\bf P}\times \hat {\bf
q}_T),
\end{equation}
where $x$ is the proton light-cone momentum fraction carried by the
gluon, $F_g(x,q_T)$ is the unpolarised TMD  parton density,
$\Delta^N F_g^{\uparrow}(x,q_T)$ is the Sivers function,
$q_T=|{\bf q}_T|$ and symbol $(\ \hat {}\ )$ denotes a unit vector, $\hat {\bf a}={\bf
a}/|{\bf a}|$.

In the present paper we are interested in accessing the gluon Sivers function using the TSSA of inclusive $J/\psi$-production. However, the task of studying gluon
distributions using charmonia is rather challenging theoretically.
Production of charmonia proceeds in two stages: first, a
$c\bar{c}$-pair is produced at short distances, predominantly
via gluon-gluon fusion but also with a non-negligible contribution
of $q\bar q-$and $qg-$initiated subprocesses. The second stage is
hadronization of $c\bar{c}$-pair into a physical charmonium
state, which proceeds essentially nonperturbatively, at large
distances (low scales) and is accompanied by a complicated
rearrangement of color via exchanges of soft gluons between the
$c\bar{c}$-pair and other colored partons produced in the
collision. At present, two approaches to describe $c\bar{c}$ hadronization are most popular: the non-relativistic QCD (NRQCD) factorization
\cite{Bodwin:1994jh} and (Improved) Color-Evaporation Model (CEM)
\cite{CEM,Ma:2016exq,Cheung:2018tvq}.

In a context of TMD-factorization, first rigorous results for heavy quarkonium physics have been obtained only very recently \cite{Echevarria:2019ynx,Fleming:2019pzj}, showing that the
TMD-factorization formula for quarkonium production will differ from
the case of Drell-Yan  pair or Higgs-boson production processes. For quarkonia, the TMD-factorization formula have to include additional shape-functions with the corresponding
evolution. However, in the present paper we adopt a simpler phenomenological approach of Generalized Parton Model (GPM) which will be described in more details in Sec.~\ref{subsec:GPM}. In the standard GPM approach the Sivers function is assumed to be process-dependent, because effects of initial (ISI) and final-state interactions (FSI) are factorized into this function. An alternative approach is Color-Gauge-Invariant (CGI) GPM formalism of Refs.~\cite{Gamberg:2011,DAlesio:2011kkm,DAlesio:2013cfy,DAlesio:2017rzj,DAlesio:2019gnu,DAlesio:2020eqo}, which we summarize in Sec.~\ref{subsec:NRQCD} and~\ref{subsec:CEM}. In this framework the process-dependent ISIs and FSIs are lifted from Sivers-like TMD PDF to the coefficient function, using the one-gluon exchange approxiamtion. Thus in CGI-GPM  results for Sivers function extracted from different processes can be directly compared.

The behavior of TSSA in the process $p^{\uparrow}p \to J/\psi X$ has been studied recently  in Refs.~\cite{DAlesio:2017rzj,DAlesio:2019gnu,Godbole:2017syo}. In
Refs.~\cite{DAlesio:2017rzj,DAlesio:2019gnu}, $J/\psi$ production
was treated in color-singlet approximation~\cite{DAlesio:2017rzj} and full NRQCD approach\cite{DAlesio:2019gnu}, including color-octet states, respectively. In case of color-singlet mechanism, the CGI-GPM
corrections to GPM cross sections have been included. In
Ref.~\cite{Godbole:2017syo}, the ICEM was used and authors
investigated the effect of the evolution of TMD PDFs involved,
on the asymmetry $A_N$.

The aim of the present study, is to to provide reasonable estimates for TSSA-effects which could be observed in the kinematic conditions of planned Spin Physics Detector experiment at NICA collider~\cite{Savin:2015paa, Arbuzov:2020cqg}. We will compare estimates for TSSA, obtained in GPM and CGI-GPM frameworks and will explore both theoretical approaches for charmonium hadronization -- NRQCD and ICEM. In case of NRQCD, we will show that at small transverse momenta of produced charmonium $k_{T{\cal C}}< m_{\cal C}$, color-singlet mechanism describes data for prompt $J/\psi$
production and inclusion of color-octet terms is not needed at LO in $\alpha_s$. Therefore for our NRQCD predictions for TSSA within GPM and CGI-GPM we include only color-singlet states of final $c\bar{c}$-pair. In connection to this findings, we comment on the results of of Refs. \cite{DAlesio:2017rzj,DAlesio:2019gnu} in Sec.~\ref{sec:Res}.

The present paper is organized as follows. In the Sec.~\ref{sec:models}, we
present basic formulas of GPM and CGI-GPM for $2\to 1$ and $2\to 2$ partonic
subprocesses, as well as details on NRQCD approach and ICEM. In the Sec.~\ref{sec:Res}, our numerical results for transverse-momentum spectra of prompt $J/\psi$ and TSSA $A_N$ in $p^{\uparrow}p \to
J/\psi X$ at PHENIX RHIC and SPD NICA are presented and discussed.
In this section we also compare our results with the similar results obtained earlier for PHENIX
RHIC at $\sqrt{s}=200$ GeV both in NRQCD approach
\cite{DAlesio:2017rzj,DAlesio:2019gnu} and ICEM
\cite{Godbole:2017syo}.

\section{Theoretical formalism}
\label{sec:models}

\subsection{GPM and CGI-GPM}
\label{subsec:GPM} The factorization scheme, which is suitable to describe inclusive observables in hadronic collisions, depends on hierarchy between typical hard scale of the process and involved transverse momenta. Collinear Parton Model (CPM) is used for studies of high-$p_T$ production, when hard
scale $(\mu \sim p_T \gg \Lambda_{QCD})$. Such a way, in CPM we neglect transverse momenta of partons initiating the hard process and hadronic cross section can be factorized as convolution of hard coefficients and collinear parton distribution
functions (PDFs) $f_{a}(x,\mu_F)$ at the factorization scale $\mu_F$ with $\mu_F \simeq \mu\sim p_T$ and $a=q,\bar q,g$. In the kinematical domain of CPM, transverse momenta of final-state particles
$({\bf k}_T)$ are generated in the hard scattering of partons and
influence of small intrinsic transverse momenta of partons in
hadrons (${\bf q}_T$), which originates from nonperturbative
effects, can be neglected. The typical estimate for average squared intrinsic
transverse-momentum of a parton is $ \langle q_T^2\rangle
\simeq 1$ GeV$^2$ or even smaller.

If one is interested in particle production with small transverse momenta $k_T\simeq \sqrt{\langle q_T^2\rangle} \ll \mu$
in a hard processes with the scale $\mu$, effects of intrinsic parton motion in
hadrons have to be taken into account. In the Transverse Momentum Dependent (TMD) factorization, the  hadronic cross section is expressed as a convolution of TMD PDFs $F_{a}(x,q_T,\mu, \zeta)$ and hard partonic
cross section. The TMD factorization theorem has been proven in the
limit of $k_T \ll \mu$ or $q_T \ll \mu$, and TMD PDFs evolve with respect to two scales: $\mu$ and $\zeta$, where the latter one is the so-called rapidity scale related with rapidity divergences~\cite{Collins:2011zzd}. The typical value of hard scale $\mu$ in charmonium ($\cal C$)
production is given by charmonium mass, $m_{\cal C}=3.1$ -- $3.7$ GeV, so TMD factorization should be used in the region $k_T\ll m_{\cal C}$. The region $k_T \simeq m_{\cal C}$ requires matching with finite-order perturbative corrections and possible nonperturbative power-suppressed corrections to the TMD term.

The Generalized Parton Model (GPM) is a simplified version of TMD factorization, which is generally applied for phenomenological estimates of various observables in proceses for which TMD factorization have not been rigorously proven yet. Typically TMD PDFs in GPM are parametrized by a simple factorized prescription:
\begin{equation}\label{eq:FactHypo}
F_a(x,q_T,\mu_F)=f_a(x,\mu_F)G_a(q_T),
\end{equation}
where $f_a(x,\mu_F)$ is corresponding collinear PDF. The dependence
on transverse-momentum of a parton is described by Gaussian distribution
 $G_a(q_T)= \exp[{-q_T^2/\langle
q_T^2\rangle_a}]/(\pi\langle q_T^2\rangle_a)$ with normalization condition $\int d^2q_T
G_a(q_T)=1$. And effects of TMD-evolution w.r.t. rapidity scale~\cite{Collins:2011zzd, Vladimirov:2019odu} $\zeta$ are neglected, which means that GPM estimates are applicable only in a narrow range of hard and rapidity scales $\zeta\sim\mu$. The latter condition is however always fulfilled in our case, since the scale of the process of charmonium production at low $k_{T{\cal C}}$ is given by $m_{\cal C}$.

Within the GPM, the differential cross section for charmonium production in proton-proton collisions for the $2\to 1$-type hard subprocess $g(q_1)+g(q_2) \to {\cal C}(k)$ can be written as
follows
\begin{equation}
d\sigma (pp\to {\cal C}X)=\int dx_1 \int d^2q_{1T} \int dx_2 \int
d^2q_{2T} F_g(x_1,{\bf q}_{1T},\mu_F) F_g(x_2,{\bf q}_{2T},\mu_F)
d\hat\sigma(gg\to {\cal C})\label{eq:2to1},
\end{equation}
where ${\cal C}=J/\psi, \psi(2S)$, or $\chi_c(1P)$, and
\begin{equation}
d\hat\sigma(gg\to {\cal C})=(2\pi)^4 \delta^{(4)}(q_1+q_2-k)
\frac{\overline{|M(gg\to {\cal C})|^2}}{2 x_1 x_2 s}
\frac{d^4k}{(2\pi)^3}\delta_+(k^2-m_{\cal C}^2)
\end{equation}
with $s=2P_1P_2$ -- the squared center of mass energy of the $pp$-collision. Integrating-out delta functions one obtains:
\begin{equation}
\frac{d\sigma (pp\to {\cal C} X)}{d^2k_T}=\frac{\pi}{s} \int
\frac{dx_1}{x_1} \int d^2q_{1T} \int \frac{dx_2}{x_2}
F_g(x_1,{\bf q}_{1T},\mu_F) F_g(x_2,{\bf q}_{2T},\mu_F) \overline{|M(gg\to {\cal
C})|^2}\delta(\hat s-m_C^2)\label{eq:2to1kt},
\end{equation}
where $\hat s=k^2=(q_1+q_2)^2$, ${\bf q}_{2T}={\bf k}_T-{\bf
q}_{1T}$, $k^\mu=(k_0,{\bf k}_T,k_z)^\mu$.
For consistency of GPM and to avoid problems with gauge invariance of hard scattering amplitudes, four-momenta of initial-state partons have to be put on mass-shell $(q_1^2=q_2^2=0)$. Hence $q_{1,2}$ read:
\begin{eqnarray}
q_1^\mu&=&\left(x_1\frac{\sqrt{s}}{2}+\frac{{\bf q}_{1T}^2}{2\sqrt{s}x_1},{\bf
q}_{1T}, x_1\frac{\sqrt{s}}{2}-\frac{{\bf q}_{1T}^2}{2\sqrt{s}x_1}
\right)^\mu , \label{eq:q1_mu} \\
q_2^\mu&=&\left(x_2\frac{\sqrt{s}}{2}+\frac{{\bf q}_{2T}^2}{2\sqrt{s}x_2},{\bf
q}_{2T}, -x_2\frac{\sqrt{s}}{2}+\frac{{\bf q}_{2T}^2}{2\sqrt{s}x_2}\right)^\mu, \label{eq:q2_mu}
\end{eqnarray}
where  $x_{1,2}$ are the proton light-cone momentum fractions carried
by the partons:
$$x_1=\frac{q_{1}^0+q_{1}^3}{\sqrt{s}},\qquad x_2=\frac{q_{2}^0-q_{2}^3}{\sqrt{s}}.$$
As follows from Eqs. (\ref{eq:q1_mu}) and (\ref{eq:q2_mu}):
\begin{eqnarray}
\hat s&=&
x_1 x_2s+2{\bf q}_{1T}^2-2|{\bf q}_{1T}| |{\bf k}_T|\cos(\phi_1)+\frac{{\bf q}_{1T}^2{\bf q}_{2T}^2}{x_1 x_2
s}.\label{eq:s2}
\end{eqnarray}
The latter result allows one to integrate-out delta function $\delta(\hat s-m_C^2)$ taking the integral over $dx_2$ in Eq.~(\ref{eq:2to1kt}). To perform the calculation of transverse-momentum
spectrum in fixed rapidity interval one inserts a Heaviside theta-function implementing the rapidity cut under the sign of integral (\ref{eq:2to1kt}), while the rapidity of $\cal C$ can be calculated as
$$y=\ln\left(\frac{k^0+k^3}{m_{{\cal C}T}} \right), \quad m_{{\cal C}T}=\sqrt{m_{\cal C}^2+k_T^2}, \quad k^0+k^3=x_1 \sqrt{s}+\frac{q_{2T}^2}{x_2 \sqrt{s}}.$$

The cross section for the
$2\to  2$ subprocess ($g(q_1)+g(q_2)\to {\cal C}(k) +g(q_3)$, ${\cal
C}=J/\psi, \psi(2S)$) is given by formula (\ref{eq:2to1}) with
\begin{equation}
d\hat\sigma(gg\to {\cal C} g)=(2\pi)^4 \delta^{(4)}(q_1+q_2-k-q_3)
\frac{\overline{|M(gg\to {\cal C} g)|^2}}{2 x_1 x_2 s}
\frac{d^3k}{(2\pi)^3 2k_0}\frac{d^4q_3}{(2\pi)^3} \delta_+(q_3^2).
\end{equation}
Replacing $q_3^2\to \hat{s}+\hat{t}+\hat{u}-m_{\cal C}^2$ with $\hat{t}=(q_1-k)^2$, $\hat{u}=(q_2-k)^2$
one can remove integral over $d^4q_3$ by delta function and obtain
\begin{eqnarray}
k_0\frac{d\sigma}{d^3k}=\frac{d\sigma}{dyd^2k_T}&=&\frac{1}{16\pi^2}\int
dx_1 \int d^2q_{1T} \int dx_2\int d^2q_{2T}F_g(x_1,{\bf q}_{1T},\mu_F)
F_g(x_2,{\bf q}_{2T},\mu_F) \times \nonumber \\
&&\times \frac{\overline{|M(gg\to {\cal C} g)|^2}}{x_1 x_2
s}\delta(\hat s+\hat t+\hat u-m_{\cal C}^2),
\end{eqnarray}
where $$\hat t=m_{\cal C}^2-\sqrt{s}m_{{\cal C}T}x_1e^{-y}-\frac{{\bf q}_{T1}^2 m_{{\cal C}T}e^y}{\sqrt{s}x_1}+2|{\bf q}_{T1}||{\bf k}_{T}|\cos(\phi_1),$$
$$\hat u=m_{\cal C}^2-\sqrt{s}m_{{\cal C}T}x_2e^{y}-\frac{{\bf q}_{T2}^2 m_{{\cal C}T}e^{-y}}{\sqrt{s}x_2}+2|{\bf q}_{T2}||{\bf k}_{T}|\cos(\phi_2).$$ Then one can take integral over $dx_2$
analytically using delta function $\delta(\hat s+\hat t+\hat
u-m_C^2)$. In such a way, no kinematic approximations is made and exact $2\to 1$ and $2\to 2$ kinematics is implemented in presence of transverse-momentum of initial-state partons.

Master formulas presented above have been used directly in calculations of
prompt $J/\psi$ production in the NRQCD approach, see Sec.~(\ref{subsec:NRQCD}). In the case of ICEM approach the treatment of $2\to 1$ subprocess is slightly different,
see Sec.~\ref{subsec:CEM}.

In this paper we study TSSAs, usually denoted by $A_N$, measured in
$p^{\uparrow}p \to {\cal C} X \quad({\cal
C}=J/\psi,\chi_c,\psi(2S))$ inclusive reactions and defined as:
\begin{equation}
A_N=\frac{d\sigma^{\uparrow}
-d\sigma^{\downarrow}}{d\sigma^{\uparrow}
+d\sigma^{\downarrow}}=\frac{d \Delta\sigma}{2d\sigma} 
\label{eq:AN},
\end{equation}
where $\uparrow,\downarrow$ are opposite proton spin orientations
perpendicular to the scattering plane in $pp$ center-of-mass frame.
The numerator and denominator of $A_N$ reads
\begin{eqnarray}
d \Delta\sigma &\propto& \int dx_1 \int d^2q_{1T} \int dx_2 \int
d^2q_{2T} \bigl[\hat F_g^{\uparrow}(x_1,{\bf q}_{1T},\mu_F)-\hat
F_g^{\downarrow}(x_1,{\bf q}_{1T}, \mu_F) \bigr] \nonumber
\\ &\times& F_g(x_2,{\bf q}_{2T},\mu_F) d\hat\sigma(gg\to {\cal C }X),\label{eq:numSSA} \\
d\sigma &\propto& \int dx_1 \int d^2q_{1T} \int dx_2 \int d^2q_{2T}
F_g(x_1,{\bf q}_{1T},\mu_F) F_g(x_2,{\bf q}_{2T},\mu_F) d\hat\sigma(gg\to {\cal C}X),\label{eq:denSSA}
\end{eqnarray}
where $\hat F_g^{\uparrow,\downarrow}(x,q_{T},\mu_F)$ is the
distribution of unpolarized gluon (or quark) in polarized proton.
 Following the Trento
conventions \cite{Bacchetta:2004jz}, the gluon Sivers function(GSF) can be
introduced as
\begin{eqnarray}
\Delta \hat F_g^{\uparrow}(x_1,{\bf q}_{1T},\mu_F)& \equiv & \hat
F_g^{(\uparrow)}(x_1,{\bf q}_{1T},\mu_F)-\hat
F_g^{(\downarrow)}(x_1,{\bf q}_{1T},\mu_F)= \nonumber
\\
&=& \Delta^N F_g^{\uparrow}(x_1,{\bf q}^2_{1T},\mu_F)\cos(\phi_1)=-2\frac{q_{1T}}{M_p}F^g_{1T}(x_1,q_{1T},\mu_F)\cos(\phi_1),
\label{eq:sivers1}
\end{eqnarray}
and GSF has to satisfy the positivity bound
\begin{eqnarray}\label{eq:Positivity}
\frac{q_{1T}}{M_p}\left|F^g_{1T}(x_1,q_{1T},\mu_F)\right|\leq F_g(x_1,q_{1T},\mu_F),
\end{eqnarray}
where $M_p$ -- mass of the proton.

We adopt factorized Gaussian parametrizations for both
the unpolarized TMD distribution $F_g(x,q_{T},\mu_F)$ and the Sivers
function $\Delta^N F_g^{\uparrow}(x,q^2_{T},\mu_F)$:
\begin{equation}
\Delta^N
F_g^{\uparrow}(x,q_{T}^2,\mu_F)=2N_g(x)F_g(x,q_T,\mu_F)h(q_T),
\end{equation}
\begin{equation}
N_g(x)=N_g x^\alpha (1-x)^\beta
\frac{(\alpha+\beta)^{\alpha+\beta}}{\alpha^\alpha \beta^\beta},
\end{equation}
\begin{equation}
h(q_T)=\sqrt{2e}\frac{q_T}{M'}e^{-q_T^2/M'^2},
\end{equation}
which satisfies the bound (\ref{eq:Positivity}) for any values of $\alpha$ and $\beta$.
After introducing the parameter
\begin{equation}
\rho_g=\frac{M'^2}{M'^2+\langle q_T^2\rangle_g}, \quad 0<\rho_g<1,
\end{equation}
we write for gluon Sivers function (GSF):
\begin{eqnarray}
\Delta^N
F_g^{\uparrow}(x,q_{T}^2,\mu_F)=2\frac{\sqrt{2e}}{\pi}N_g(x)f_g(x,\mu_F)\sqrt{\frac{1-\rho_g}{\rho_g}}\frac{q_T}{\langle
q_T^2\rangle_g^{3/2} }e^{-q_T^2/\rho_g \langle
q_T^2\rangle_g}.\label{eq:sivers-fin}
\end{eqnarray}
In our numerical calculations we will use two different of GSF obtained earlier in Refs.
\cite{DAlesio:2015fwo} which we call SIDIS1 for brevity, and \cite{DAlesio:2018rnv} which we refer to as GSF parametrization by D'Alesio
{\it et al.}. Corresponding values of parameters are collected in the Table \ref{table:1}.

\begin{table}[h]
\begin{tabular}{|c|c|c|c|c|c|}
\hline
 GSF set & $N_g$ & $\alpha_g$ & $\beta_g$ & $\rho_g$ & $\langle
q_T^2\rangle_g$, GeV$^2$   \\ \hline SIDIS1 & 0.65& 2.8& 2.8& 0.687
& 0.25  \\ \hline
D'Alesio {\it et al.} & 0.25 & 0.6& 0.6& 0.1& 1.0 \\
\hline
\end{tabular}
\caption{Parameters of GSFs} \label{table:1}
\end{table}

  To introduce the CGI-GPM let us first recall the explanation of Sivers effect, which had been described for the first time in Ref.~\cite{Brodsky:2002cx}. In this paper it was shown, that Sivers asymmetry in semi-inclusive DIS (SDIS) process at leading twist is a quantum effect generated by exchanges of soft gluons between initial (ISI) or final-state (FSI) partons produced in a hard process, and spectator system originating as a remnant of an incoming hadron. In standard TMD factorization, this soft gluons are taken into account within the gauge-invariant definition of Sivers-like TMD PDF, which contains Wilson lines. The sign of Sivers TMD PDF depends on the direction of Wilson lines, which can be past- or future-pointing, representing the ``space-time trajectory'' of an initial-state or struck quark produced respectively in Drell-Yan or SDIS hard-scattering process. Thus the Sivers function in standard TMD and perhaps in GPM approaches is process-dependent and it is not clear how to extend factorization for Sivers effect to the processes with colored final-states, like $J/\psi$ production.

\begin{figure}[h]
\begin{center}
\begin{tabular}{|c|c|}
\hline
\parbox[c]{0.2\textwidth}{\includegraphics[width=0.2\textwidth, trim = 0cm 0.6cm 1cm 0cm]{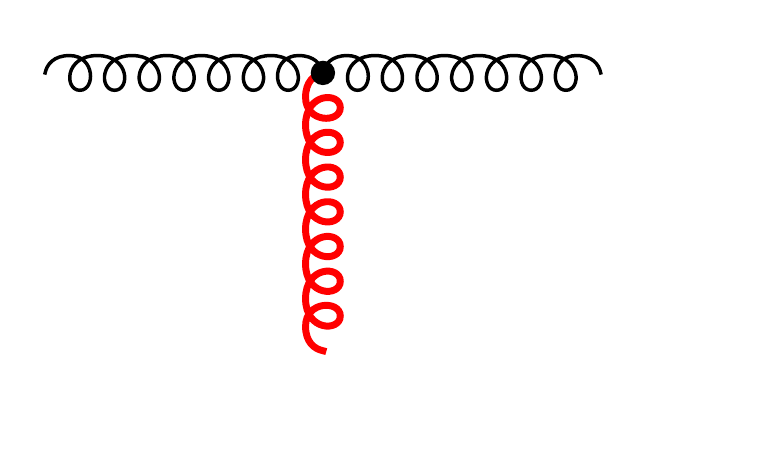}} & \parbox[c]{0.2\textwidth}{$T^a_{cb}$} \\
\hline
\parbox[c]{0.2\textwidth}{\includegraphics[width=0.2\textwidth, trim = 0cm 0.6cm 1cm 0cm]{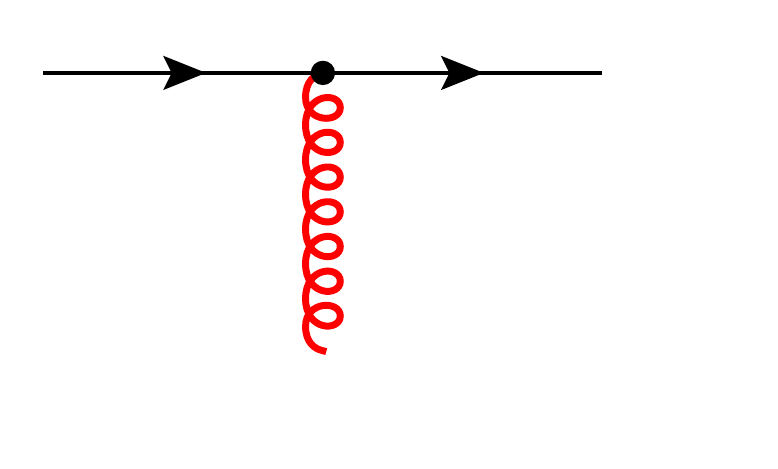}} & $t^c_{ji}$ \\
\hline
\parbox[c]{0.2\textwidth}{\includegraphics[width=0.2\textwidth, trim = 0cm 0.6cm 1cm 0cm]{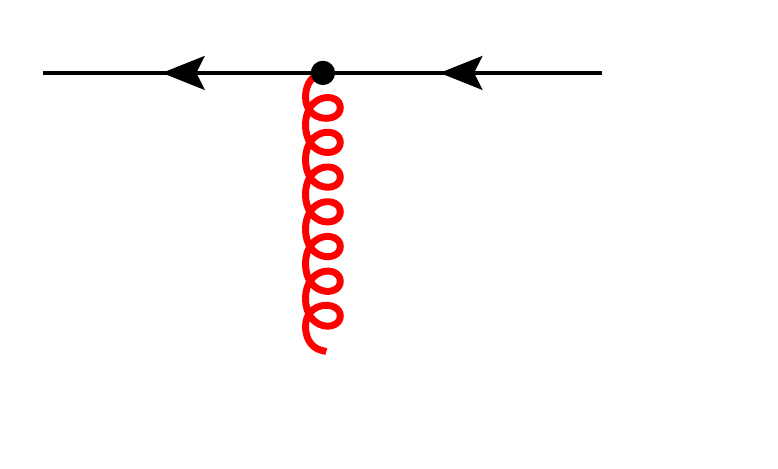}} & -$t^c_{ij}$ \\
\hline
\parbox[c]{0.2\textwidth}{\includegraphics[width=0.2\textwidth]{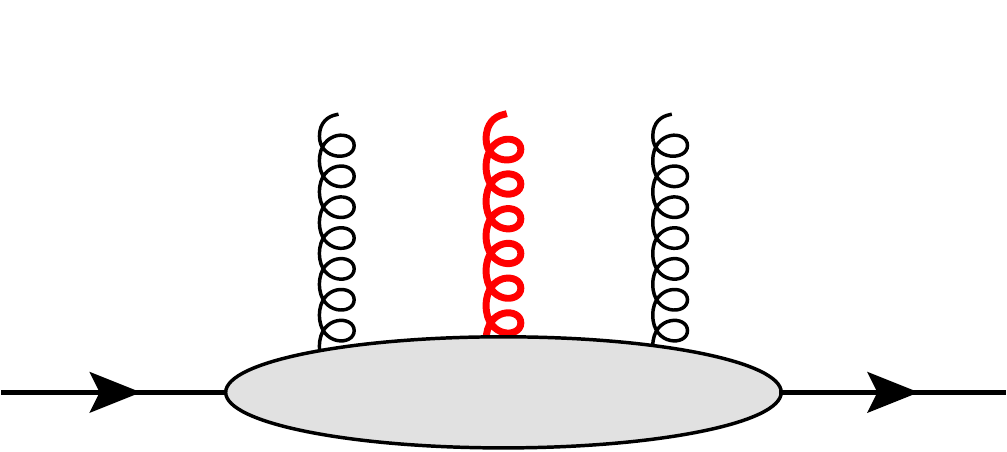}} & \parbox[c]{0.2\textwidth}{$\mathcal{T}^c_{ab}$ -- for $f$-type\\ $\mathcal{D}^c_{ab}$ -- for $d$-type\\} \\
\hline
\parbox[c]{0.2\textwidth}{\includegraphics[width=0.2\textwidth]{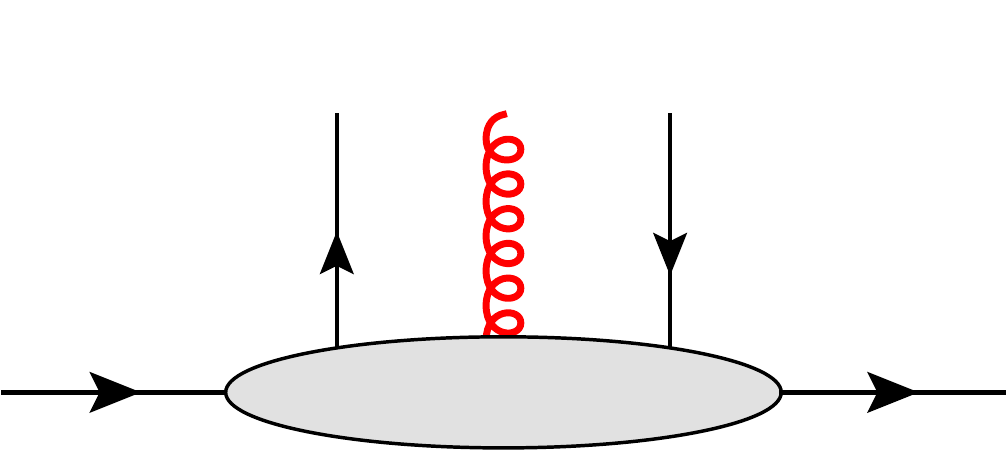}} & \parbox[c]{0.2\textwidth}{$\mathcal{Q}^a_{ij}$} \\
\hline
\end{tabular}
\end{center}
\caption{\label{fig:FRs}
Feynman rules for color factors within the CGI-GPM with additional eikonal gluon.}
\end{figure}

  The aim of CGI-GPM~\cite{Gamberg:2011,DAlesio:2011kkm,DAlesio:2013cfy,DAlesio:2017rzj,DAlesio:2019gnu,DAlesio:2020eqo} formalism is to extract above-mentioned process-dependence from the TMD PDF to the hard-scattering coefficient. The effects of ISI and FSI are included in CGI-GPM via one-gluon exchange approximation~\cite{Gamberg:2011,DAlesio:2017rzj}.  For the case of gluon Sievers effect, this approximation leads to appearance of independent GSFs of $f$-type ($F_{1T}^{g(f)}$) and $d$-type ($F_{1T}^{g(d)}$) corresponding to two independent ways of combining three gluons into a color-singlet (Fig.~\ref{fig:FRs}). The coupling of additional ``eikonal'' gluon from the GSF to the hard process leads only to modification of the color structure of the latter one. There is no four-momentum transfer from the additional gluon to the hard process, because Sivers effect comes from imaginary part of the loop integral over momentum of exchanged gluon~\cite{Brodsky:2002cx}, while the latter one is saturated by the contribution of the soft region. Moreover, only coupling of eikonal gluon to the initial-state or {\it observed} colored final-state particles contributes to the asymmetry, while effects of coupling to un-observed final-state partons cancel-out between amplitude and complex-conjugate amplitude. While arguments above are specific for the one-gluon exchange approximation, an additional argument in favor of CGI-GPM is, that it's hard-scattering coefficients reproduce coefficients in the twist-3 collinear approach~\cite{Gamberg:2011}.

  In the Fig.~\ref{fig:FRs} we collect the corresponding Feynman rules and prescriptions for calculation of the hard scattering coefficient for the numerator~(\ref{eq:numSSA}) of the TSSA within CGI GPM. For detailed derivation see e.g., Ref.~\cite{DAlesio:2017rzj}. The color projectors for $f$ and $d$-type GSFs are defined as follows:
\begin{equation}\label{eq:ColProj}
\mathcal{T}^c_{ab}=\mathcal{N_T}(-if_{abc}),\quad \mathcal{D}^c_{ab}=\mathcal{N_D}d_{abc},\quad \mathcal{Q}^a_{ij}=\mathcal{N_Q}t^a_{ij},
\end{equation}
where $f_{abc}(d_{abc})$ is totally antisymmetric (symmetric) structure constant of the $SU(N_c)$ color gauge group, $t^a_{ij}$ -- the generators of $SU(N_c)$ group in the fundamental representation, and
\begin{equation}\label{eq:ColNorm}
\mathcal{N_T}=\frac{1}{N_c(N_c^2-1)},\quad \mathcal{N_D}=\frac{N_c}{(N_c^2-4)(N_c^2-1)},\quad \mathcal{N_Q}=\frac{2}{N_c(N_c^2-1)}
\end{equation}
with $N_c=3$.

 In the following sections we mostly use the CGI-GPM hard-scattering coefficients which already had been calculated by other authors, in which case we cite the corresponding reference. However in Sec.~\ref{subsec:CEM} we will need new (to our knowledge) CGI-GPM hard-scattering coefficient for $gg\to c\bar c$-process with {\it both} final-state $c$-quarks being observed.

\subsection{NRQCD}
\label{subsec:NRQCD} In the framework of the NRQCD-factorization
approach, the cross section of charmonium production via a
partonic subprocess $a + b \to {\cal C} + X$ is given by a double expansion in powers of $\alpha_s$ and squared relative velocity of heavy quarks in a bound state $v^2$ as:
\begin{equation}
d\hat \sigma (a + b \to {\cal C} + X)=\sum_n d\hat \sigma (a + b \to
c\bar c[n] + X)\langle{\cal O}^{\cal C}[n]\rangle,
\end{equation}
where $n$ denotes the set of color, spin, orbital and total angular
momentum quantum numbers of the $c\bar c$ pair and the four-momentum
of the latter is assumed to be equal to the one of the physical
quarkonium state ${\cal C}$. The cross section $d\hat \sigma (a + b
\to c\bar c[n] + X)$ can be calculated in perturbative QCD as an
expansion in $\alpha_s$. The nonperturbative transition of the $c\bar c$ pair into ${\cal C}$ is
described by the Long-Distance Matrix Elements (LDMEs) $\langle
{\cal O}^{\cal C}[n]\rangle$. Color-singlet LDMEs can be determined from measured decay widths of charmonia using the known next-to-leading-order (NLO) QCD result or  from calculations in potential models \cite{Eichten:1995ch}. Color-octet LDMEs are
considered as free parameters in charmonium production
cross sections.  Typically LDMEs up to NNLO ($O(v^4)$) in $v^2$-scaling are included in NRQCD-factorization calculations: $n = {^3S}_1^{(1)}, {^3S}_1^{(8)}, {^1S}_0^{(8)},
{^3P}_J^{(8)}$ if ${\cal C} = J/\psi, \psi^\prime$ and $n =
{^3P}_J^{(1)}, {^3S}_1^{(8)}$ if ${\cal C} = \chi_{cJ}$, where
$J=0,1,2$. For the LDMEs of $P$-vawe states we adopt the known Heavy-Quark Spin Symmetry relations, which are valid up to $O(v^2)$:
\begin{eqnarray}
\langle{\cal O}^{\psi(nS)}[^3P_J^{(1,8)}]\rangle&=&(2J+1)\langle{\cal
O}^{\psi(nS)}[^3P_0^{(1,8)}]\rangle,\nonumber\\
\langle{\cal O}^{\chi_{cJ}}[^3P_J^{(1)}]\rangle&=&(2J+1)\langle{\cal
O}^{\chi_{c0}}[^3P_0^{(1)}]\rangle,\nonumber\\
\langle{\cal O}^{\chi_{cJ}}[^3S_1^{(8)}]\rangle&=&(2J+1)\langle{\cal
O}^{\chi_{c0}}[^3S_1^{(8)}]\rangle.
\end{eqnarray}

 From general point of view, LDMEs should be universal and
process-independent parameters. However in practice their numerical values strongly
depend on approach which is used to describe $c\bar c-$pair
production and data included into the fit. For example one can compare color-octet LDMEs obtained in LO CPM \cite{Cho:1995ce,Cho:1995vh}, in NLO CPM\cite{Butenschoen:2011yh} and in the $k_T-$factorization approach
\cite{Kniehl:2006sk,Saleev:2012hi}. Neevertheless, the hierarchy expected from velocity-scaling rules: $\langle{\cal O}^{\cal C}[^3P_0^{(1)}]\rangle \gg \langle{\cal O}^{\cal C}[^3P_0^{(8)}]\rangle$, $\langle{\cal O}^{\cal C}[^3S_1^{(1)}]\rangle \gg \langle{\cal O}^{\cal C}[^3P_J^{(8)}]\rangle \gg  \left( \langle{\cal O}^{\cal C}[^3S_1^{(8)}]\rangle, \langle{\cal O}^{\cal C}[^1S_0^{(8)}]\rangle \right)$ is respected by all the fits.

We adopt the following values of color-singlet LDMEs \cite{Barbieri:1981xz}: $\langle{\cal
O}^{J/\psi}[^3S_1^{(1)}]\rangle = 1.3$~GeV$^3$, $\langle{\cal
O}^{\psi^\prime}[^3S_1^{(1)}]\rangle = 6.5\times10^{-1}$~GeV$^3$,
and $\langle{\cal O}^{\chi_{cJ}}[^3P_J^{(1)}]\rangle = (2 J +
1)\times 8.9\times 10^{-2}$~GeV$^5$.

Squared LO in $\alpha_S$ amplitudes for $2\to 1$ subprocesses  in CPM are well-known \cite{Cho:1995vh}:
\begin{eqnarray}
\overline{|{\cal A}(g + g \to {\cal C}[^{3}P_0^{(1)}]|^2}
&=&\frac{8}{3}\pi^2\alpha_s^2 \frac{\langle{\cal O}^{\cal
C}[^3P_0^{(1)}]\rangle}{M^3},\label{eq:3P01}
\\
\overline{|{\cal A}(g + g \to {\cal C}[^{3}P_1^{(1)}]|^2}
&=&0,\label{eq:3P11}
\\
\overline{|{\cal A}(g + g \to {\cal C}[^{3}P_2^{(1)}]|^2}
&=&\frac{32}{45}\pi^2\alpha_s^2 \frac{\langle{\cal O}^{\cal
C}[^3P_2^{(1)}]\rangle}{M^3},\label{eq:3P21}
\\
\overline{|{\cal A}(g + g \to {\cal C}[^{3}S_1^{(8)}]|^2}
&=&0,\label{eq:3S18}
\\
\overline{|{\cal A}(g + g \to {\cal C}[^{1}S_0^{(8)}]|^2}
&=&\frac{5}{12}\pi^2\alpha_s^2 \frac{\langle{\cal O}^{\cal
C}[^1S_0^{(8)}]\rangle}{M},\label{eq:3S08}
\\
\overline{|{\cal A}(g + g \to {\cal C}[^{3}P_0^{(8)}]|^2}
&=&5\pi^2\alpha_s^2\frac{\langle{\cal O}^{\cal
C}[^3P_0^{(8)}]\rangle}{M^3},\label{eq:3P08}
\\
\overline{|{\cal A}(g + g \to {\cal C}[^{3}P_1^{(8)}]|^2}
&=&0,\label{eq:3P18}
\\
\overline{|{\cal A}(g + g \to {\cal C}[^{3}P_2^{(8)}]|^2}
&=&\frac{4}{3}\pi^2\alpha_s^2 \frac{\langle{\cal O}^{\cal
C}[^3P_2^{(8)}]\rangle}{M^3},\label{eq:3P28}
\\
\overline{|{\cal A}(q + \bar q \to {\cal C}[^{3}S_1^{(8)}]|^2}
&=&\frac{16}{27}\pi^2\alpha_s^2 \frac{\langle{\cal O}^{\cal
C}[^3S_1^{(8)}]\rangle}{M}.\label{eq:3S18q}
\end{eqnarray}

As it was discussed above, in GPM these subprocesses contribute to transverse-momentum spectrum because TMD PDFs are involved.

There is only one relevant LO in $\alpha_S$ $2\to 2$ partonic
subprocess,  which describes direct production of $J/\psi$
or $\psi(2S)$ via color-singlet intermediate state $[^{3}S_1^{(1)}]$
as it is in the Color-Singlet Model, because $2\to 2$ subprocesses producing color-octet states of $c\bar{c}$-pair are of NLO in $\alpha_s$ in the GPM approach for $k_{T{\cal C}}$-spectrum.  The squared amplitude for this partonic
subprocess reads~\cite{Gastmans:1986qv}:
\begin{eqnarray}
\overline{|{\cal A}(g + g \to {\cal C}[^{3}S_1^{(1)}] + g|^2}
&=&\pi^3 \alpha_s^3 \frac{\langle{\cal O}^{\cal
C}[^3S_1^{(1)}]\rangle}{M^3}\, \frac{320 M^4}{81 (M^2 - {\hat t})^2
(M^2 - {\hat u})^2 ({\hat t} + {\hat u})^2}
\nonumber\\
&&{}\times (M^4 {\hat t}^2 - 2 M^2 {\hat t}^3 + {\hat t}^4 + M^4
{\hat t} {\hat u} - 3 M^2 {\hat t}^2 {\hat u} + 2 {\hat t}^3 {\hat
u} + M^4 {\hat u}^2
\nonumber\\
&&{} - 3 M^2 {\hat t} {\hat u}^2 + 3 {\hat t}^2 {\hat u}^2 - 2 M^2
{\hat u}^3 + 2 {\hat t} {\hat u}^3 + {\hat u}^4).\label{eq:3S11}
\end{eqnarray}

\begin{figure}[h]
\begin{center}
\includegraphics[width=0.33\textwidth]{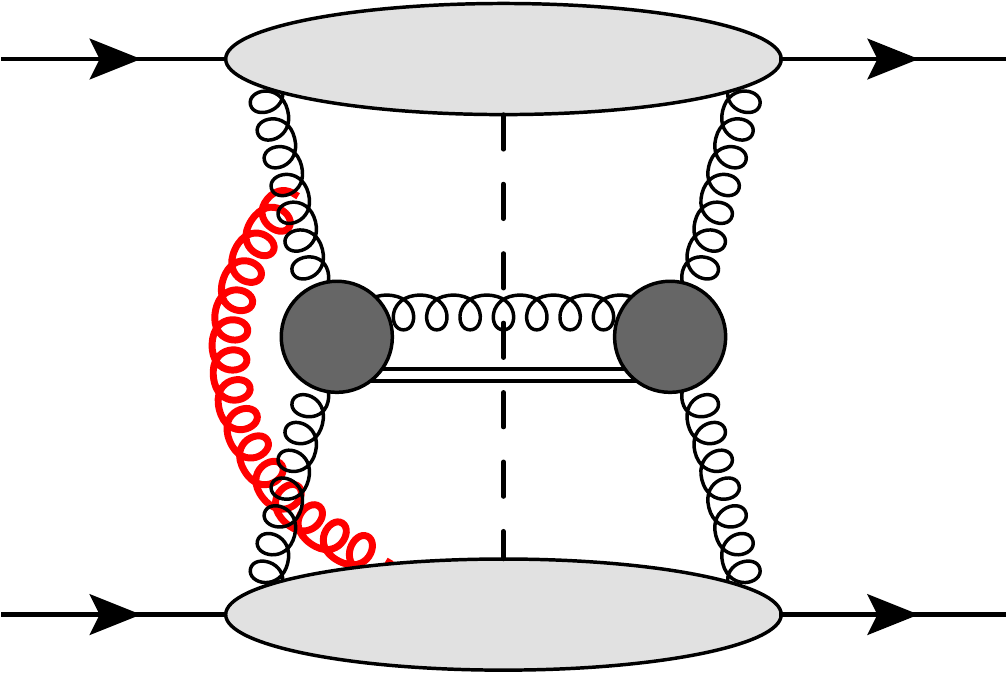}\hfill\includegraphics[width=0.33\textwidth]{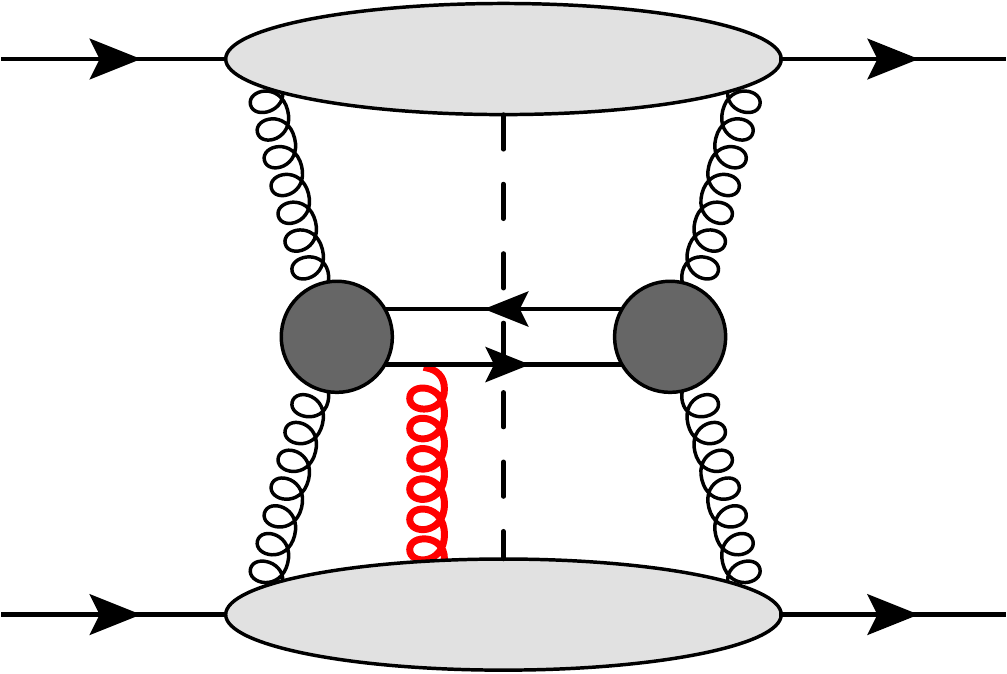}\hfill\includegraphics[width=0.33\textwidth]{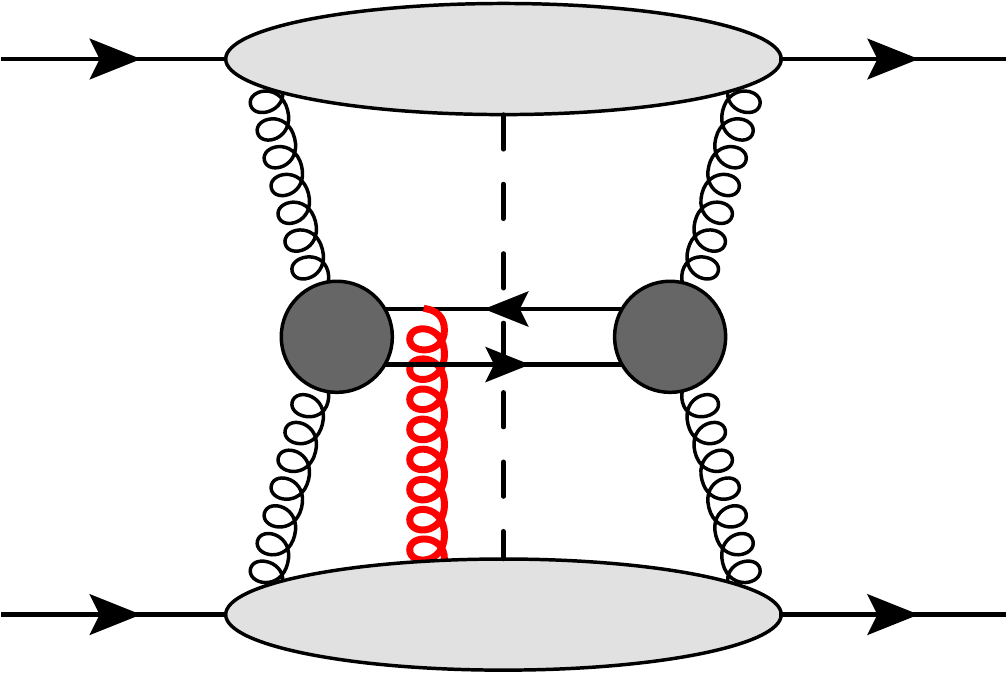}
\end{center}
\caption{\label{fig:CGI-MEs}
Example diagrams for contributions to the numerator of TSSA in CGI-GPM. Left panel: ISI for production of $^3S_1^{(1)}$-state, middle and right panels: FSI for $gg\to c\bar{c}$-process with both final-state quarks tagged}
\end{figure}

 Turning now to the case of CGI-GPM factorization we introduce the following notations: $\overline{|\mathcal{A}|^2_{GPM}}$ for above-described matrix elements of a hard subprocess in the GPM and $H^{(f/d)}_{CGI}$ for the coefficient-function, obtained within the CGI-GPM factorization prescription. Then, following Ref.~\cite{DAlesio:2017rzj}, one writes-down the contribution of the $2\to 1$ or $2\to 2$ subprocess of production of $^3P_J^{(1)}$ or $^3S_1^{(1)}$-states of $c\bar{c}$-pair to the numerator of TSSA as:
\begin{multline}\label{eq:CGIsubst}
F_{1T}^{g(f)}\otimes H^{(f)}_{CGI}+F_{1T}^{g(d)}\otimes H^{(d)}_{CGI} = \\ =\frac{C_I^{(f)}+C_{F_c}^{(f)}}{C_U}F_{1T}^{g(f)}\otimes \overline{|\mathcal{A}|^2_{GPM}}+
\frac{C_I^{(d)}+C_{F_c}^{(d)}}{C_U}F_{1T}^{g(d)}\otimes \overline{|\mathcal{A}|^2_{GPM}},
\end{multline}
where $F_{1T}^{g(f)}$ and $F_{1T}^{g(d)}$ are above-mentioned $f$-type ($C$-even) and $d$-type ($C$-odd) GSFs, and $\otimes$ denotes convolution in the light-cone momentum fraction and transverse-momentum of gluon from the polarized proton. Here $C_U$ is the color factor of the unpolarized cross section, which corresponds to the usual QCD result, while $C_I^{(f/d)}$ and $C_{F_c}^{(f/d)}$ are modified color factors corresponding to ISI and FSI in CGI GPM. In case of color-singlet state of $c\bar{c}$-pair only ISIs (first diagram in the Fig.~\ref{fig:CGI-MEs}) contribute in both cases ${}^3S_1^{(1)}$ and ${}^3P_J^{(1)}$, so that $C_{F_c}^{(f)}=C_{F_c}^{(d)}=0$,  while as it was shown in Refs.~\cite{Gamberg:2011,DAlesio:2011kkm,DAlesio:2013cfy,DAlesio:2017rzj,DAlesio:2019gnu,DAlesio:2020eqo}:
\begin{equation}\label{eq:ISIcolfact}
C_I^{(f)}=-\frac{1}{2}C_U,\qquad C_I^{(d)}=0,
\end{equation}
for the case of ${}^3S_1^{(1)}$ final-state ($2\to 2$ process), while
\begin{equation}\label{eq:ISIcolfact_P}
C_I^{(f)}=C_U,\qquad C_I^{(d)}=0,
\end{equation}
for ${}^3P_J^{(1)}$ states ($2\to 1$ processes). One notices, that in both cases, $d$-type GSF does not contribute, so that only $f$-type GSF is relevant for color-singlet model.

 Such a way, accounting for the effects of ISI and FSI with CGI-GPM formalism leads to a smaller numerical values of TSSA in charmonium hadroproduction within color-singlet approximation of NRQCD-factorization, as compared to the ordinary GPM.

A comment on the treatment of color-octet contributions in the CGI-GPM approach of Refs.~\cite{DAlesio:2019gnu,DAlesio:2020eqo} is in order here. As it is explained in Ref.~\cite{DAlesio:2020eqo} around Eq. (A34) the coefficient functions contributing to the numerator of the asymmetry formula for $2\to 1$-subprocesses producing color-octet states of $c\bar{c}$-pair in this approach are equal to zero due to cancellation between ISI and FSI contributions, unlike coefficient functions of ordinary GPM in our Eqs. (\ref{eq:3S08}), (\ref{eq:3P08}) and (\ref{eq:3P28}). Due to this fact, authors of Refs.~\cite{DAlesio:2019gnu,DAlesio:2020eqo} had to include contributions of $2\to 2$-processes to obtain non-zero effect on the asymmetry from color-octet channels. However, since LO in $\alpha_s$ contribution in CGI-GPM is zero, one expects that NLO $2\to 2$ contribution should not contain any infra-red or collinear divergences. Contrary to this expectation, e.g. coefficient function for the contribution of $g+g\to c\bar{c}\left[ {}^1S_0^{(8)} \right]+g$-subprocess to the numerator of the asymmetry (Eq. (A3) of Ref.~\cite{DAlesio:2020eqo}) clearly contains non-integrable singularities at $\hat{t}\to 0$ or $\hat{s}\to m_{\cal C}^2$ which due to non-zero transverse-momentum of initial-state partons will lead to divergent cross section in GPM. This fact has been mentioned in Ref.~\cite{DAlesio:2019gnu}, where it is admitted that a regulator $\mu_{\rm IR}\sim 0.8$ GeV for this divergences have to be introduced. In our opinion, appearance of non-regulated divergences in CGI-GPM deserves further study and we are going to address this problem in the future. In the numerical calculations of the present paper we include only color-singlet NRQCD channels, which are free from above-mentioned problem. We should also point-out, that usage of color-octed LDMEs of NRQCD, obtained in the NLO fits together with LO coefficient functions is not consistent, because NLO corrections to short-distance coefficients in NRQCD are very significant and even change sign of $P$-wave contributions. Nevertheless, such NLO LDMEs had been used in Refs.~\cite{DAlesio:2019gnu,DAlesio:2020eqo}. The good agreement of un-polarized cross section calculated in full NRQCD with NLO LDMEs in Ref.~\cite{DAlesio:2019gnu} with experimental data is probably due to neglecting of the feed-down contributions by these authors, while this contributions are actually non-negligible (see our Tab.~\ref{Tab:DirFD}).  Also, the treatment of $2\to 1$  kinematics in Refs.~\cite{DAlesio:2019gnu,DAlesio:2020eqo} is different from ours, which leads to significant numerical effects for NICA energies.

To calculate feed-down contribution in prompt $J/\psi$ production we
use following branching ratios which are taken from
Ref.~\cite{Eidelman:2004wy}: $B(\psi^\prime \to J/\psi + X)=0.576$, $B(\chi_{c0} \to J/\psi +
\gamma)=0.012$, $B(\chi_{c1} \to J/\psi + \gamma)=0.318$ and
$B(\chi_{c2} \to J/\psi + \gamma)=0.203$, while $B(J/\psi \to \mu^+ + \mu^-)=0.0601$.

\subsection{ICEM}
\label{subsec:CEM}

Main physical assumption of the ICEM is that all $c\bar c-$pairs with invariant masses below the $D\bar D$-threshold hadronize to charmonia with some probability, which is independent from angular momentum and spin quantum numbers of the $c\bar{c}$-pair. In the ICEM~\cite{Ma:2016exq,Cheung:2018tvq} the invariant mass of the intermediate charm quark-antiquark pair is constrained to be larger than the mass
of produced charmonium state, $m_{\cal C}$, instead of using the same lower
limit of integration -- $2m_c$ as it was done in the traditional CEM~\cite{CEM}. As a result, the ICEM
describes the charmonium yields as well as the ratio of $\psi(2S)$
over $J/\psi$ better than the old CEM. The partonic cross section, differential in $c\bar{c}$-invariant mass is related to the well-known total cross section of production of $c\bar{c}$-pairs as a function of partonic squared center-of-mass energy $\hat{s}$:
\begin{eqnarray}
\hat\sigma(\hat s,gg\to c\bar c)&=&\frac{\pi \alpha_S^2}{3\hat
s}\biggl[ (1+w+\frac{w^2}{16})\ln
\biggl(\frac{1+\sqrt{1-w}}{1-\sqrt{1-w}} \biggr)-\biggl(
\frac{7}{4}+\frac{31}{16}w \biggr)\sqrt{1-w} \biggr],\label{eq:totCSGPM} \\
\hat\sigma(\hat s,q \bar q\to c\bar c)&=&\frac{8 \pi \alpha_S^2}{27
\hat s} \biggl(1+\frac{w}{2}\biggr)\sqrt{1-w},
\nonumber
\end{eqnarray}
with $w=4m_c^2/\hat s$, as follows:
\begin{equation}
\frac{d\hat \sigma^{c\bar c}}{dM^2}=\hat\sigma(\hat s, ab\to c\bar
c)\delta(\hat s-M^2)\label{eq:ggcc},
\end{equation}
so that the GPM-factorization formula for production of $c\bar{c}$-pairs with invariant mass $M$ and total three-momentum ${\bf k}$ via gluon-gluon fusion can be written as:
\begin{eqnarray}
\frac{d\sigma^{c\bar c}}{dM^2 d^3{\bf k}}&=&\int dx_1 \int d^2q_{1T} \int
dx_2 \int d^2 q_{2T}\ F_g(x_1, q_{1T}, \mu_F) F_g(x_2, q_{2T}, \mu_F)
\hat\sigma(\hat s, gg\to c\bar c)\nonumber \\
&\times& \delta(\hat s-M^2)\delta^{(3)}({\bf q}_1+{\bf q}_2-{\bf
k}),\label{eq:ppcc}
\end{eqnarray}
where the invariant $\hat s=k^2=(q_1+q_2)^2$ can be represented as in Eq.
(\ref{eq:s2}). Finally, for differential cross section of chamonium
${\cal C}$ production in proton-proton collision in ICEM one has:
\begin{eqnarray}
\frac{d\sigma^{\cal C}}{d^3k}=F_C\times \int_{m_{\cal C}^2}^{4m_D^2} dM^2
\frac{d\sigma^{c\bar c}}{dM^2 d^3k},\label{eq:ppC}
\end{eqnarray}
where $F_{\cal C}$ is the process-independent hadronization probability to the charmonium state ${\cal C}$. Then one integrates-out ${\bf q}_{2T}$ and $M^2$ using delta
functions to find
\begin{eqnarray}
\frac{d\sigma^{\cal C}}{d^3k}&=&F_{\cal C}\times \int dx_1 \int
d^2q_{1T} \int dx_2\ F_g(x_1, q_{1T}, \mu_F) F_g(x_2, q_{2T}, \mu_F)
\hat\sigma(\hat s, gg\to c\bar c)\nonumber \\
&\times & \delta(q_{1}^3+q_{2}^3-k^3)\left[\theta(\hat
s-m_{\cal C}^2)-\theta(\hat s-4m_D^2) \right].\label{eq:ppC2}
\end{eqnarray}
In this equation the integral over $dx_2$ also can be removed by delta
function $\delta(q_{1}^3+q_{2}^3-k^3)$ thus obtaining the master formula
for numerical calculations. The quark-antiquark
annihilation channel for $c\bar c-$pair production has been
incorporated into the calculation in a similar way.

  In case of CGI-GPM factorization, the numerator of the TSSA~(\ref{eq:numSSA}), reads:
\begin{equation}\label{eq:CGIsubstICEM}
F_{1T}^{g(f)}\otimes \hat\sigma^{(f)}_{CGI}(\hat s,gg\to c\bar c)+F_{1T}^{g(d)}\otimes \hat\sigma^{(d)}_{CGI}(\hat s,gg\to c\bar c),
\end{equation}
where $\hat\sigma^{(f/d)}_{CGI}(\hat s,gg\to c\bar c)$ is the $f/d$-type coefficient function of CGI-GPM integrated over phase-space of final-state $c\bar{c}$-pair with fixed invariant mass $\hat{s}$ as in Eq.~(\ref{eq:totCSGPM}). Since in the ICEM both $c$- and $\bar c$-quarks in the final state are observed, the corresponding hard-scattering coefficient is different from the coefficient for $D$-meson TSSA, given e.g. in Ref.~\cite{Pisano:2019fsg}. To obtain new coefficient functions we take into account interactions of eikonal gluon with the initial-state gluon coming from un-polarized proton as well as with final-state $\bar{c}$ and $c$-quarks (middle and right diagrams of the Fig.~\ref{fig:CGI-MEs}). The $f/d$-type hard-scattering coefficients thus obtained have the form:
\begin{eqnarray*}
H_{CGI}^{(f)} (gg\to c\bar{c}) &=& \frac{8 \pi ^2 \alpha_s^2}{N_c \left(N_c^2-1\right) \tilde{t}^2
   \tilde{u}^2} \left(4 m_c^4 (\tilde{t}+\tilde{u})^2+4 m_c^2
   \tilde{t} \tilde{u} (\tilde{t}+\tilde{u})-\tilde{t} \tilde{u}
   \left(\tilde{t}^2+\tilde{u}^2\right)\right),\\
H_{CGI}^{(d)} (gg\to c\bar{c}) &=& N_c\frac{\tilde{t}-\tilde{u}}{\hat{s}}H_{CGI}^{(f)} (gg\to c\bar{c}) ,\\
H_{CGI} (q\bar{q}\to c\bar{c}) &=& -H_{CGI} (\bar{q}q\to c\bar{c})=\frac{8\pi^2\alpha_s^2 (N_c^2+1)}{\hat{s}^2 N_c^2} \left( 2m_c^2\hat{s} + \tilde{t}^2 + \tilde{u}^2 \right),
\end{eqnarray*}
where $\tilde{t}=\hat{t}-m_c^2$ and $\tilde{u}=\hat{u}-m_c^2$. Integrating this coefficient-functions over the phase-space of the final-state with fixed $c\bar{c}$ invariant-mass $\hat{s}$ one obtains:
\begin{eqnarray}\label{eq:totCSCGI}
\hat\sigma^{(f)}_{CGI}(\hat s,gg\to c\bar c)&=&\frac{\pi \alpha_S^2}{48\hat
s}\biggl[(\frac{w^2}{2}-w-1)\ln
\biggl(\frac{1-w/2+\sqrt{1-w}}{1-w/2-\sqrt{1-w}} \biggr)+2(
1+w)\sqrt{1-w} \biggr],\quad\\
\hat\sigma^{(d)}_{CGI}(\hat s,gg\to c\bar c)&=&0,\\
\hat\sigma_{CGI}(\hat s,q\bar q\to c\bar c)&=&\frac{10 \pi \alpha_S^2}{27
\hat s} \biggl(1+\frac{w}{2}\biggr)\sqrt{1-w}.
\end{eqnarray}
It is interesting, that integrated hard-scattering coefficient for $d$-type GSF is equal to zero similarly to the case of NRQCD, so that in both of our models, heavy-quarkonium TSSA is sensitive only to $f$-type GSF.

To obtain prompt-$J/\psi$ production spectra we take into account direct as well as feed-down contributions from decays of $\chi_{cJ}$ and $\psi(2S)$-states. At the stage of numerical calculation in ICEM, we put $m_c=1.2$ GeV
and charmonium masses are taken from PDG tables: $m_{J/\psi}=3.096$ GeV,
$m_{\psi(2S)}=3.686$ GeV, $m_{\chi_{c0}}=3.415$ GeV,
$m_{\chi_{c1}}=3.510$ GeV, and $m_{\chi_{c2}}=3.556$ GeV.

\section{Numerical Results}
\label{sec:Res}

\subsection{PHENIX RHIC}
\label{subsec:phenix}

To begin with, we compare theoretical predictions obtained in NRQCD-factorization
approach with recent experimental data for transverse-momentum
spectra of prompt $J/\psi$-mesons, measured by PHENIX RHIC experiment
\cite{Adare:2011vq}. In our NRQCD calculations we take the charm quark
mass as one half of mass of the physical charmonium state, $m_c=m_{\cal C}/2$, while in the ICEM calculations it is kept fixed at $m_c=1.2$ GeV. Also, in
case of feed-down production, the kinematic effect of the mass
splittings between charmonium states turns out to be significant and
 we take into account momentum-shift between high-mass charmonium state and final $J/\psi$
meson as it was done e.g. in Ref.\cite{Kniehl:2016sap}: $k_{T J/\psi}=(m_{J/\psi}/m_{\cal C}) k_{T{\cal C}}$.

Phenomenological analysis of intrinsic transverse-momentum of partons in
proton in LO and NLO of CPM~\cite{Wong:1998pq} demonstrates that for
gluon one has $\langle q_T^2\rangle \simeq 1$ GeV$^2$ and the same
estimation was obtained for $J/\psi$ production in GPM
\cite{DAlesio:2017rzj}.

Throughout our analysis the renormalization and factorization scales
has been identified and chosen to be $\mu_F=\mu_R=\xi \sqrt{k_T^2+m_{\cal C}^2}$
where $\xi$ is varied between $\xi=1/2$ and $\xi=2$ about its
default value $\xi=1$ to estimate the theoretical uncertainty due to
the freedom in the choice of scales. The resulting errors are
indicated as shaded bands in our figures, however they mostly cancel-out in asymmetries $A_N$.

\begin{figure}[h]
\begin{center}
\includegraphics[width=0.45\textwidth, clip=]{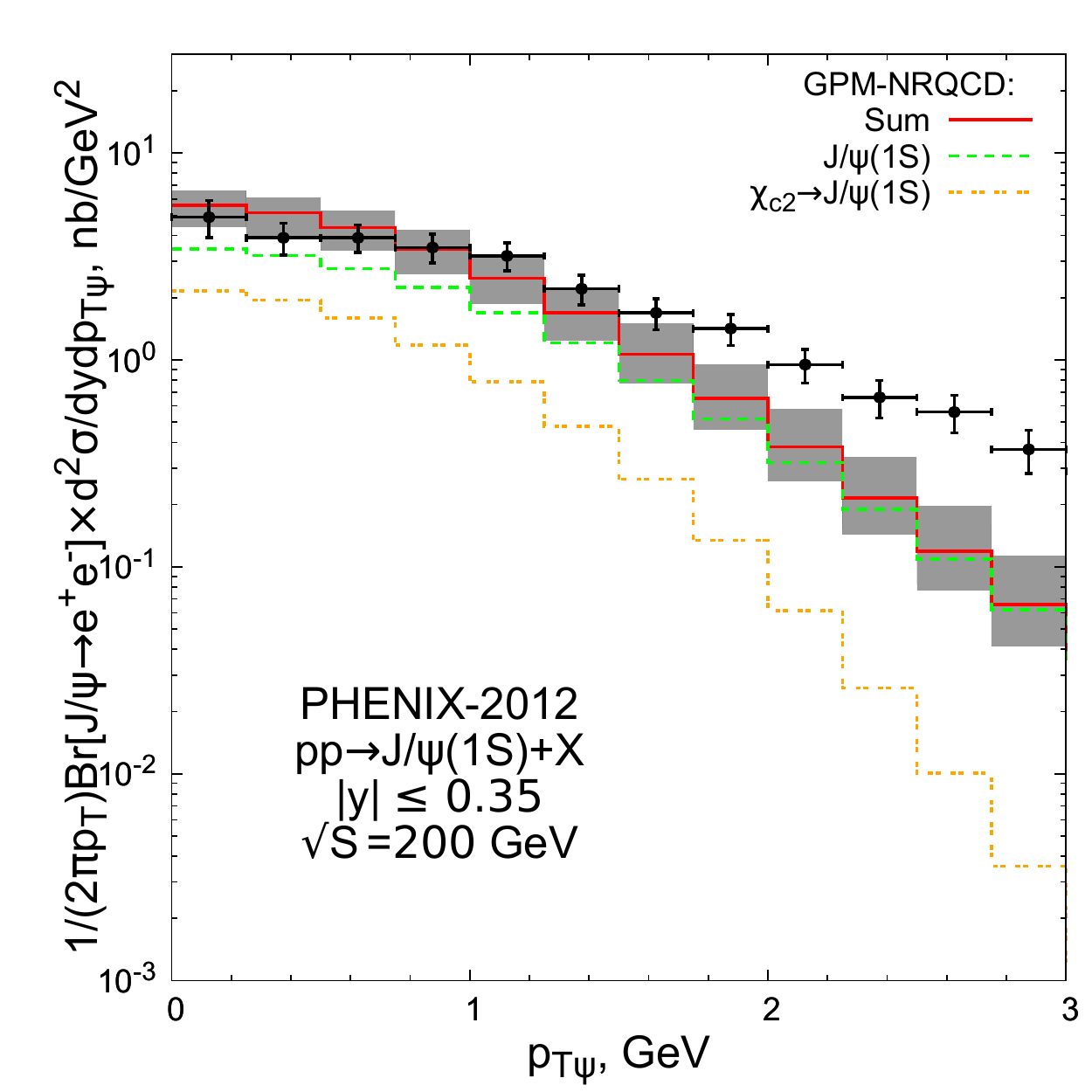}
\includegraphics[width=0.45\textwidth, clip=]{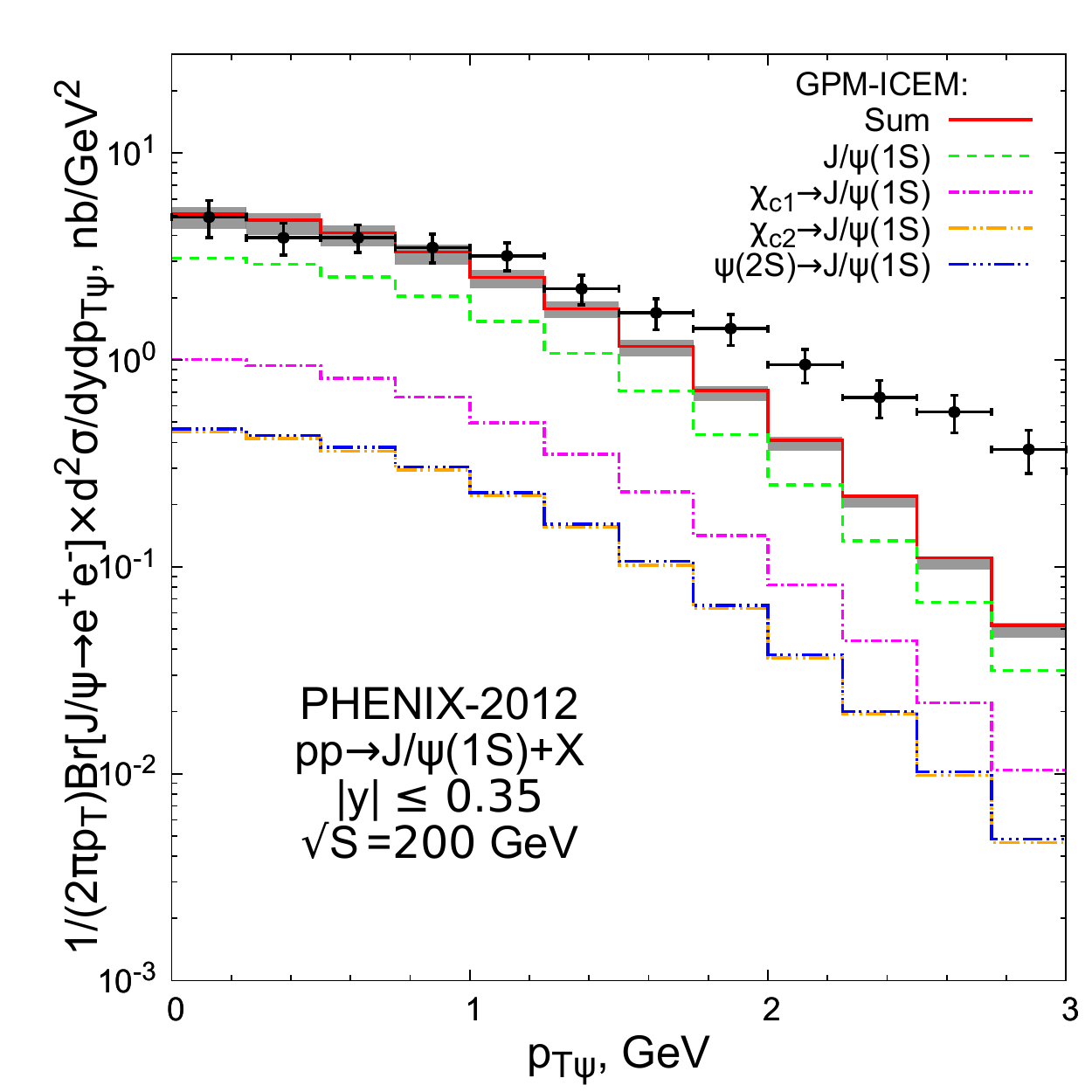}
\end{center}
\caption{\label{fig:PHENIXpt} Differential
cross section of prompt $J/\psi$ production as function of transverse-momentum at $\sqrt{s}=200$
GeV, $|y|<0.35$. The theoretical results are obtained in GPM with
$\langle q_T^2\rangle=1$ GeV$^2$. Left panel: NRQCD-factorization prediction with only color-singlet channels included. Right panel: ICEM-prediction. In the left panel, non-zero contributions from
 decays $\chi_{c0,1}\to J/\psi$ and
$\psi(2S)\to J/\psi$ are not shown.
Experimental data are from the Ref.~\cite{Adare:2011vq}.}
\end{figure}

The direct production of $J/\psi$-mesons at the $O(v^0)$ includes only contributions from CSM (\ref{eq:3S11}). The color-octet states ${}^3P_0^{(8)}$ (\ref{eq:3P08}) and ${}^3P_2^{(8)}$ (\ref{eq:3P28}) contribute at $O(v^2)$.
Intermediate state ${}^3P_1^{(8)}$ does not contribute if initial-state
partons are on-mass shell. Further color-octet contributions
(\ref{eq:3S08}),(\ref{eq:3S18}), and (\ref{eq:3S18q}) are suppressed as
$O(v^4)$ so it is natural to expect them to be negligible at small $k_{T{\cal C}}<m_{\cal C}$. In fact, similarly to the results of Ref.~\cite{DAlesio:2017rzj}, we have found that taking into account only color-singlet production mechanism, the good description of prompt $J/\psi$ transverse-momentum spectra at
small $k_{T J/\psi}< m_{J/\psi}$, i.e. in the region of
applicability of TMD-factorization, can be achieved in GPM see the left panel of Fig.~\ref{fig:PHENIXpt}. Also, our NRQCD calculation leads to  total cross section ratios of direct and feed-down contributions in good agreement with experimental data of Ref.~\cite{Adare:2011vq}, see Tab.~\ref{Tab:DirFD}.

\begin{table}[h!]
\caption{The relative contributions of direct and feed-down production within NRQCD and ICEM. Experimetal data of the PHENIX collaboration for $\sqrt{s}=200$~GeV are from~\cite{Adare:2011vq}.}\label{Tab:DirFD}
\center
\begin{tabular}{|c|c|c|}
  \hline
  $\sqrt{s}$ & Model/Source of data & $\sigma^{\rm direct}:\sigma^{\chi_c\to J/\psi}:\sigma^{\psi(2S)\to J/\psi}$ \\\hline
  \multirow{2}{*}{24 GeV}  & NRQCD & $0.58:0.39:0.03$ \\\cline{2-3}
                           & ICEM  & $0.68:0.25:0.07$ \\\hline
  \multirow{2}{*}{200 GeV} & NRQCD & $0.61:0.34:0.05$ \\\cline{2-3}
                           & ICEM  & $0.61:0.30:0.09$ \\\hline
           200 GeV         & PHENIX collab. & $0.58:0.32:0.10$ \\
  \hline
\end{tabular}
\end{table}

However, we should emphasize, that our calculations are different from calculations of the Ref.~\cite{DAlesio:2017rzj} in a respect, that we consistently take into account feed-down contributions from $\psi(2S)$ and $\chi_{cJ}$-states, while in this reference they where added very crudely, by multiplication of direct cross section by a factor $\simeq 1.4$. Such treatment of feed-down is not consistent with Color Singlet Model, since the direct $J/\psi$ and $\psi(2S)$-mesons are produced in $2\to 2$ processes in this model, while $\chi_{cJ}$-mesons are produced in $2\to 1$ processes with significantly different $k_{T{\cal C}}$-behaviour. As one can see from the left panel of Fig.~\ref{fig:PHENIXpt}, feed-down subprocesses contribute mainly at small transverse momenta in this model. Furthermore, to describe data at large transverse-momentum $k_{T J/\psi}$, $O(k_{T J/\psi}/m_{J/\psi})$-power corrections, are generated
in hard scattering by emission of additional partons, and inclusion
of color-octet contributions in direct charmonium production is absolutely necessary.

Very similar predictions for transverse-momentum spectrum can be obtained in the framework of ICEM, see the right panel of Fig.~\ref{fig:PHENIXpt}. The values of hadronization probabilities used are $F_{J/\psi}=0.02$, $F_{\chi_{c1}}=F_{\chi_{c2}}=0.06$ and $F_{\psi(2S)}=0.08$. They had been obtained via the fit of total cross section of $J/\psi$-production at PHENIX and above-mentioned experimentally-measured fractions of $J/\psi$-feeddown contribution form $\psi(2S)$ and $\chi_{cJ}$-decays (Tab.~\ref{Tab:DirFD}). This values of hadronisation probabilities are numerically close to the values obtained in Ref.~\cite{Cheung:2018tvq} by the fit of LHC data in the $k_T$-factorization approach for $c\bar{c}$-pair production.

\begin{figure}[h]
\begin{center}
\includegraphics[width=0.45\textwidth, clip=]{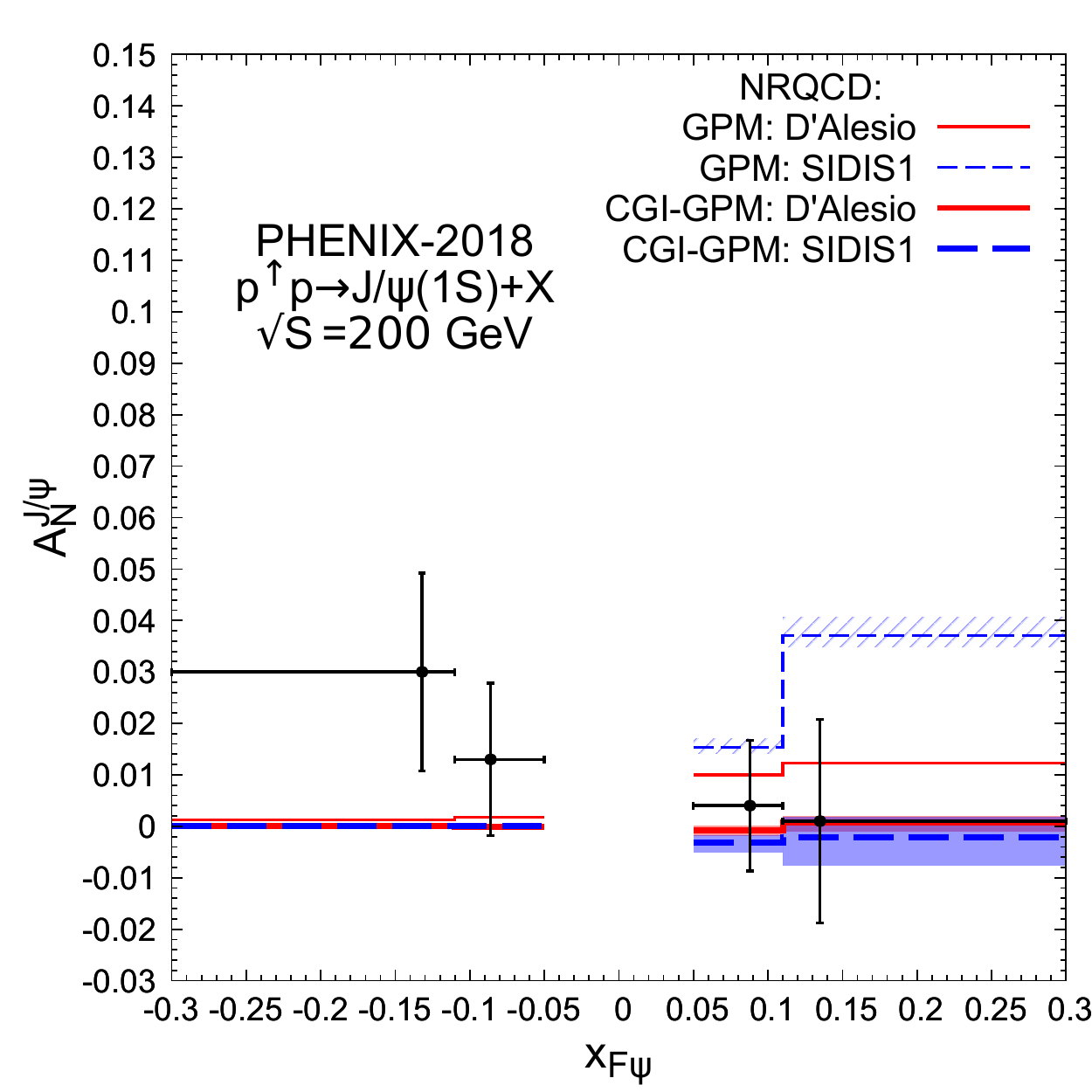}
\includegraphics[width=0.45\textwidth, clip=]{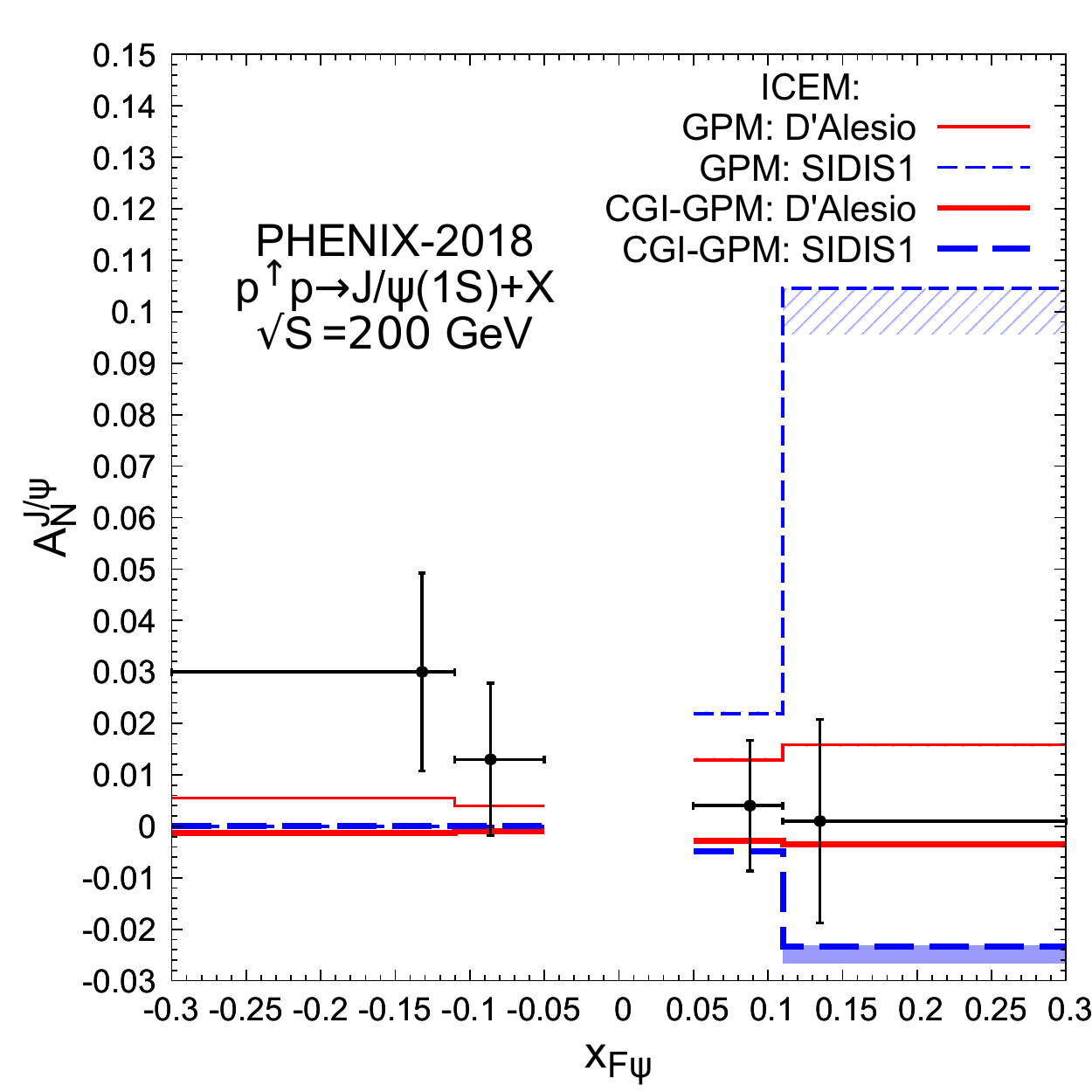}
\end{center}
\caption{\label{fig:AN-XF} TSSA $A_N^{J/\psi}$  as function of $x_F$
at $\sqrt{s}=200$ GeV within the GPM (thin histograms) and CGI-GPM(thick histograms). The theoretical results are obtained with SIDIS1 (dashed histograms) and D'Alesio {\it et al.} (solid histograms) parameterizations of GSFs. Experimental data are from Ref.~\cite{Aidala:2018gmp}. Left panel: NRQCD final-state factorization. Right panel: ICEM final-state factorization.}
\end{figure}

\begin{figure}[h]
\begin{center}
\includegraphics[width=0.45\textwidth, clip=]{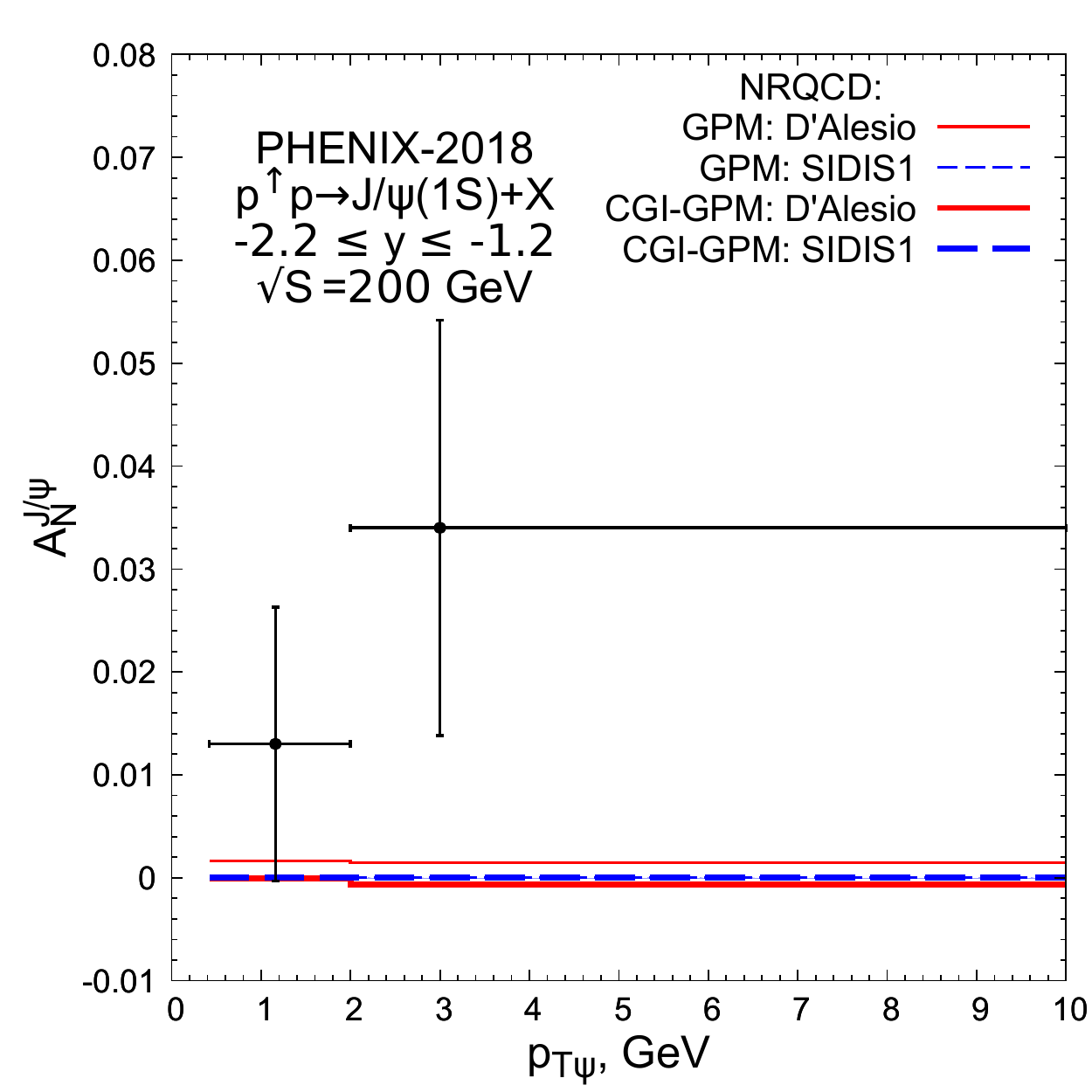}
\includegraphics[width=0.45\textwidth, clip=]{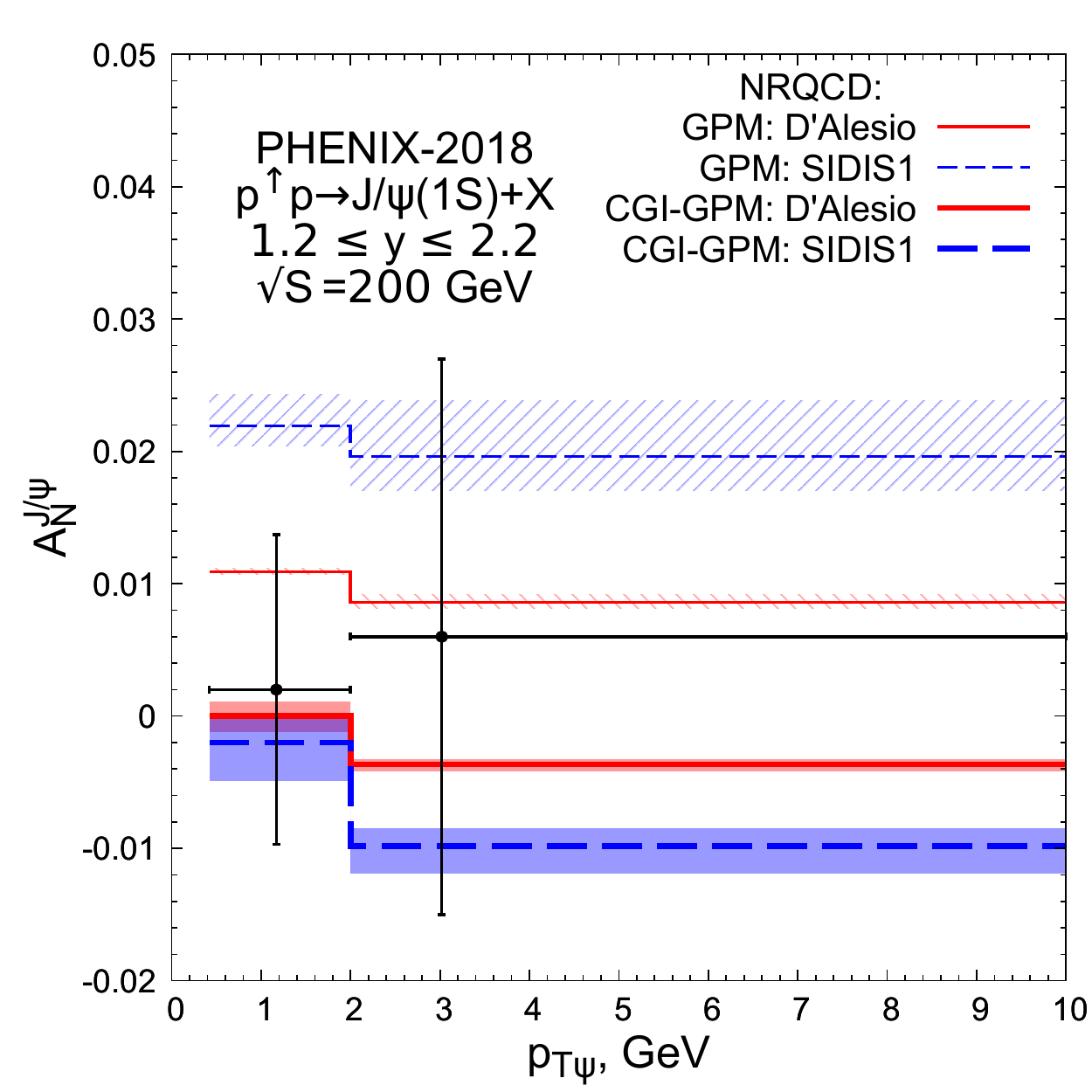}
\end{center}
\caption{\label{fig:AN-PT-NRQCD} NRQCD predictions for TSSA $A_N^{J/\psi}$ within the GPM (thin histograms) and CGI-GPM(thick histograms) as function of
$J/\psi$-transverse-momentum at $\sqrt{s}=200$ GeV. The theoretical results
are obtained with SIDIS1 (dashed lines) and D'Alesio et al. (solid lines) parameterizations of
GSFs. Left panel -- backward production $(-2.2<y<-1.2)$, right panel -- forward production $(1.2<y<2.2)$. Experimental data are from Ref.~\cite{Aidala:2018gmp}.}
\end{figure}

\begin{figure}[h]
\begin{center}
\includegraphics[width=0.45\textwidth, clip=]{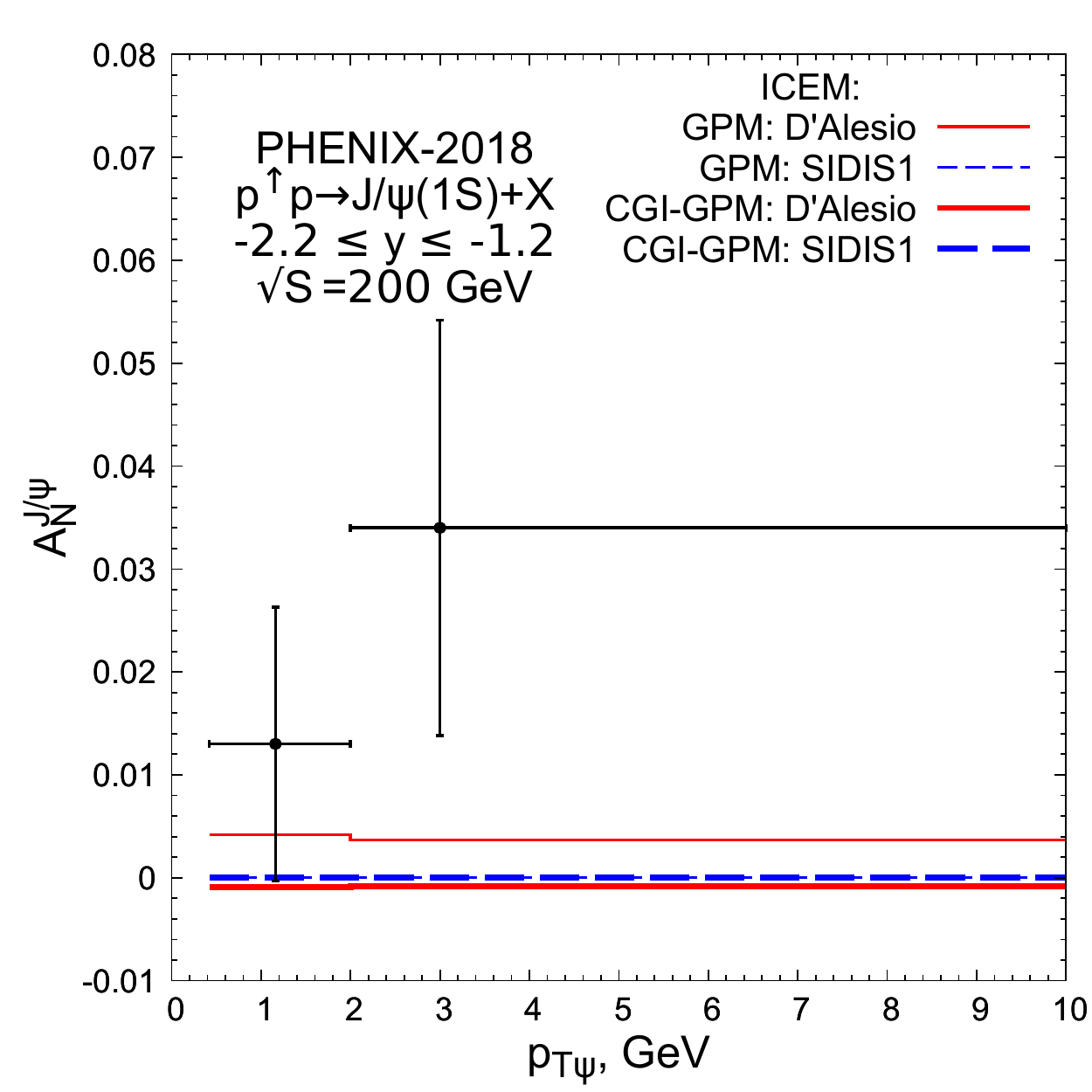}
\includegraphics[width=0.45\textwidth, clip=]{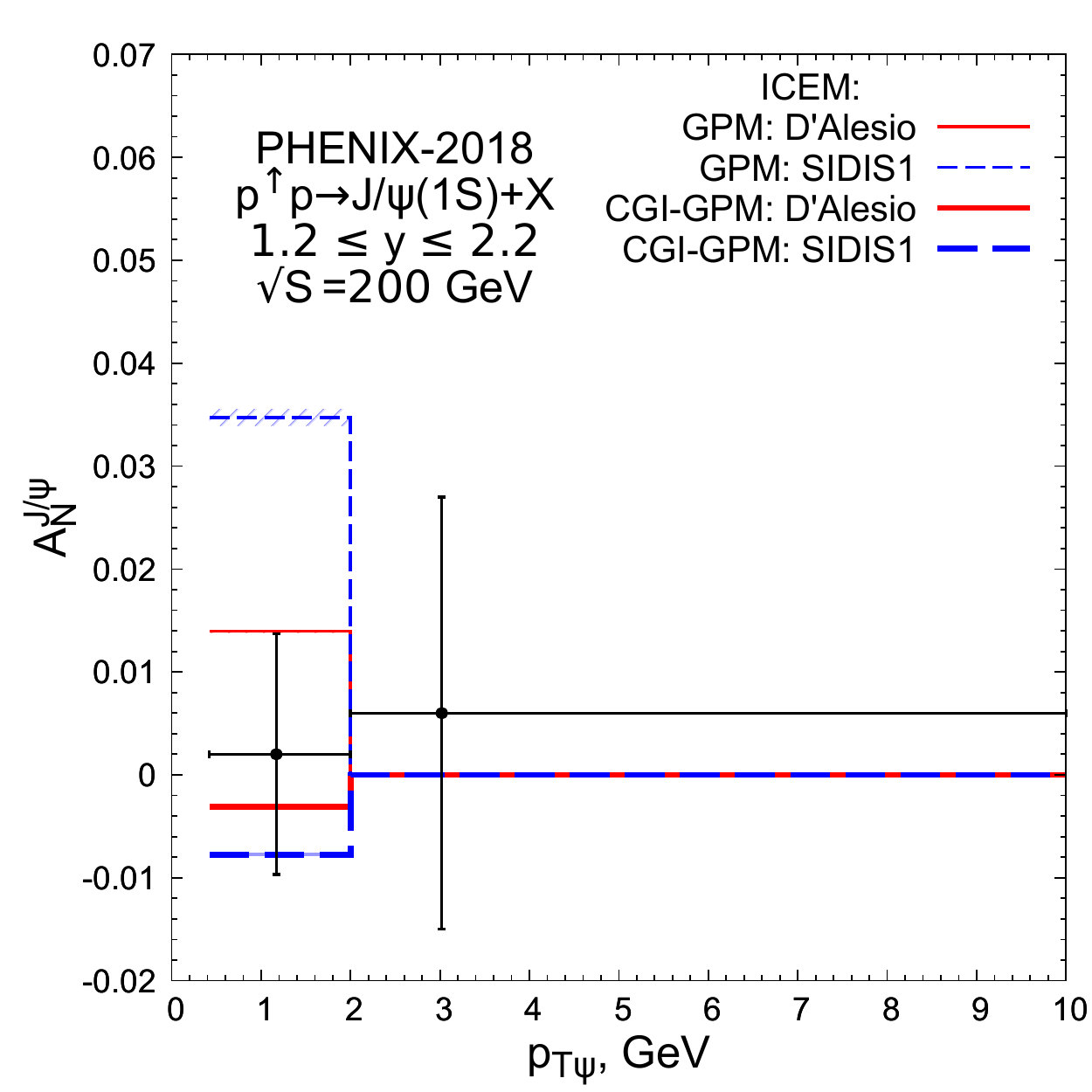}
\end{center}
\caption{\label{fig:AN-PT-ICEM} ICEM predictions for TSSA $A_N^{J/\psi}$ within the GPM (thin histograms) and CGI-GPM(thick histograms) as function of
$J/\psi$-transverse-momentum at $\sqrt{s}=200$ GeV. The theoretical results
are obtained with SIDIS1 (dashed lines) and D'Alesio et al. (solid lines) parameterizations of
GSFs. Left panel -- backward production $(-2.2<y<-1.2)$, right panel -- forward production $(1.2<y<2.2)$. Experimental data are from Ref.~\cite{Aidala:2018gmp}.}
\end{figure}

Our estimations for TSSAs at PHENIX kinematic conditions, obtained in the GPM accompanied by NRQCD-factorization approach or ICEM, are shown by thin histograms in the Fig.~\ref{fig:AN-XF} and Figs.~\ref{fig:AN-PT-NRQCD}-\ref{fig:AN-PT-ICEM} as
functions of $x_F$ and transverse-momentum respectively, together with the recent experimental data from Ref.~\cite{Aidala:2018gmp}. We conclude that within standard GPM initial-state factorization, the parametrisation for Sivers function by D'Alesio {\it et al.} is marginally consistent with experimental data for both hadronization models, while SDIS1-parametrisation predicts too large effects at positive $x_{F\psi}$ and is essentially ruled-out for the case of ordinary GPM initial-state factorization.

 The TSSA results for the CGI-GPM initial-state factorization are presented in the same Figs.~\ref{fig:AN-XF}, \ref{fig:AN-PT-NRQCD} and \ref{fig:AN-PT-ICEM} by the thick histograms. One can see that discrepancy between predictions of the CGI-GPM with the SDIS1 parametrization and experimental data is significantly reduced, rendering it to be reasonably consistent with experimental data. Another feature of CGI-GPM, evident from Figs.~\ref{fig:AN-XF}, \ref{fig:AN-PT-NRQCD} and \ref{fig:AN-PT-ICEM} is the change of sign of TSSA predicted in CGI-GPM relatively to ordinary GPM.

\subsection{SPD NICA}
\label{subsec:spd}

In this section we present our predictions for $J/\psi$ transverse-momentum spectrum and TSSA in the kinematic conditions
 of planned SPD NICA experiment in proton-proton collisions with $\sqrt{s}=24$ GeV. The SPD is expected to be an almost
  $4\pi$-geometry detector~\cite{Savin:2015paa, Arbuzov:2020cqg, Abazov:2021hku}, thus a relatively wide coverage in rapidity $|y|<3$ can be achieved.

\begin{figure}[h]
\begin{center}
\includegraphics[width=0.45\textwidth, clip=]{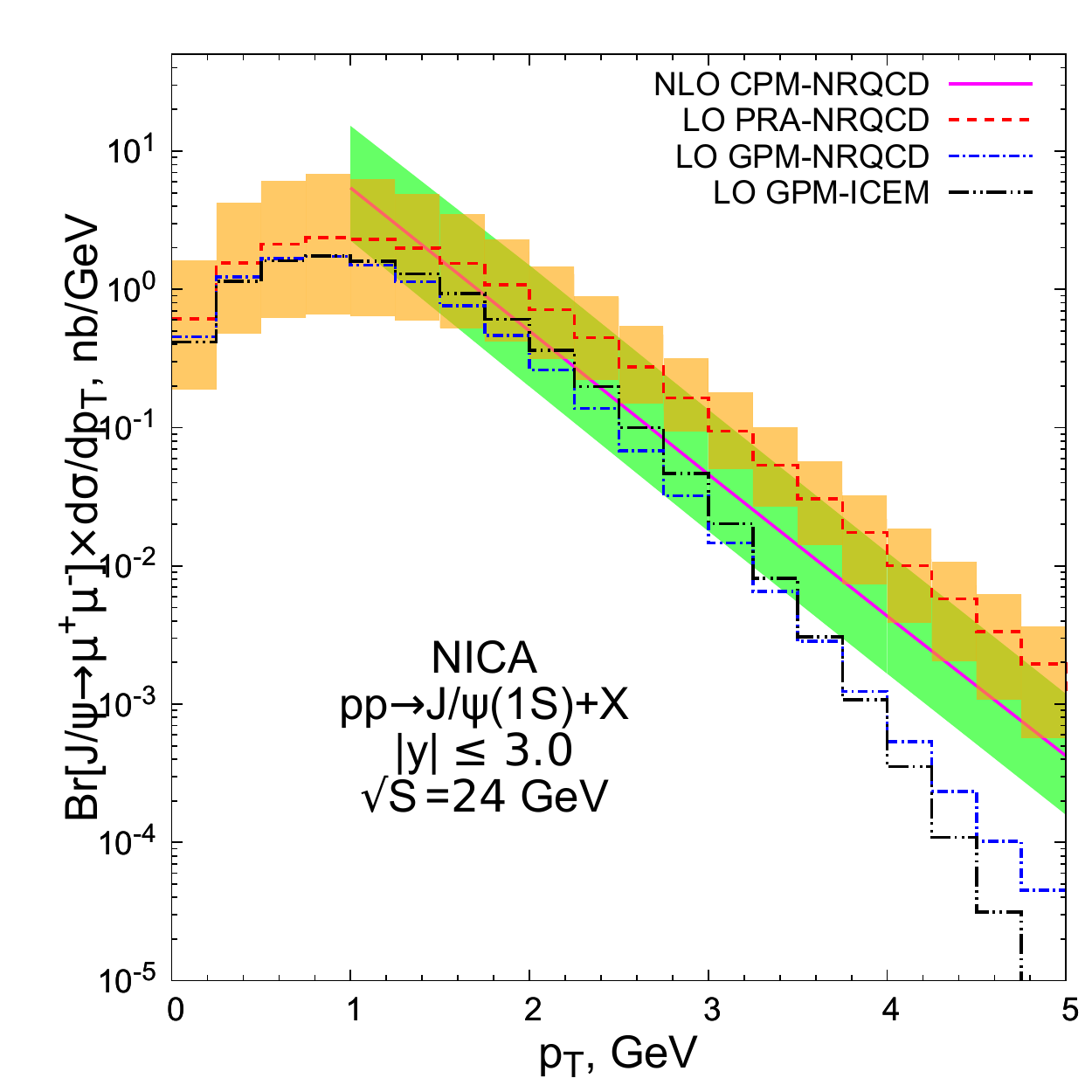}
\includegraphics[width=0.45\textwidth, clip=]{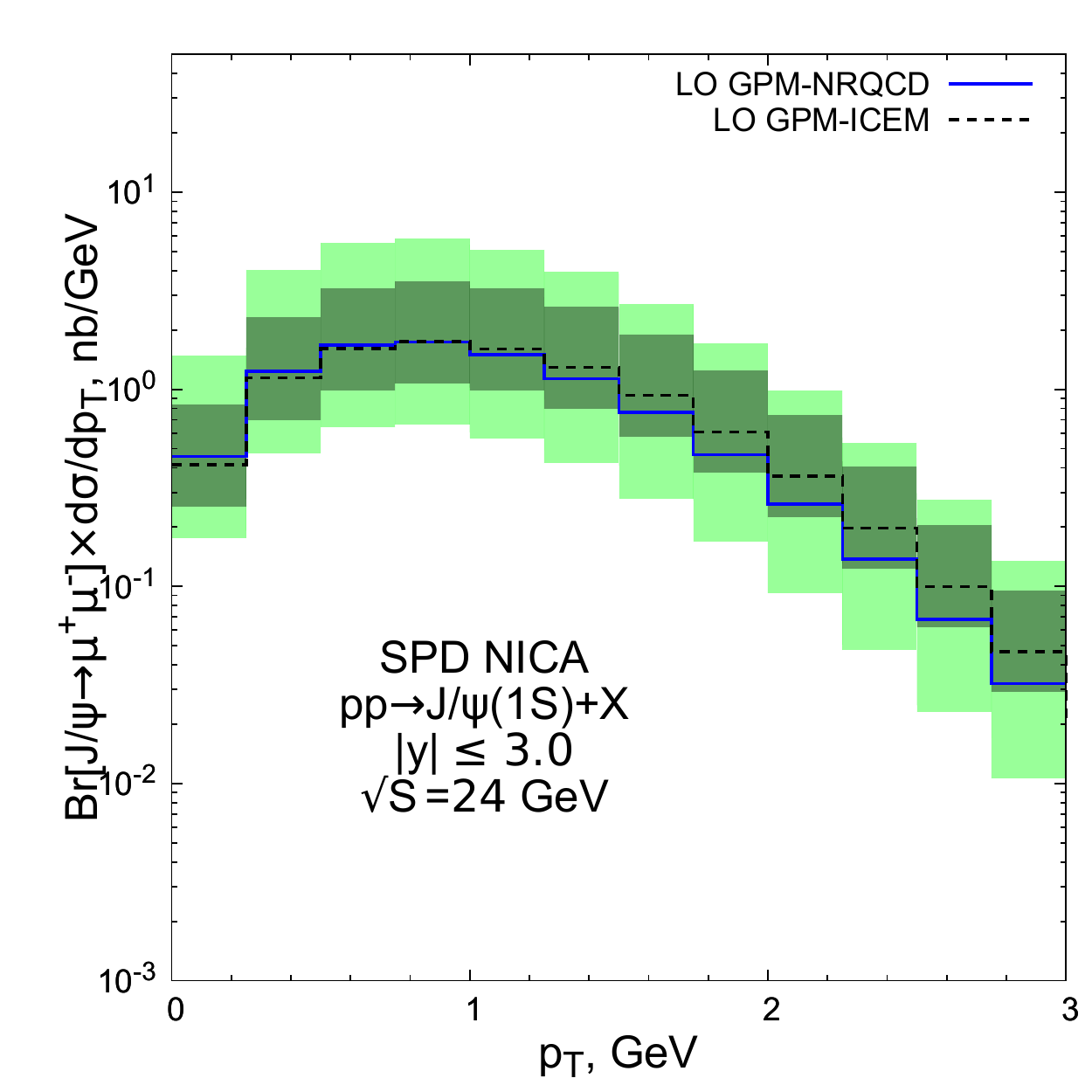}
\end{center}
\caption{\label{fig:NICApt} Prompt $J/\psi$ differential
cross section as function of $J/\psi$-transverse-momentum at $\sqrt{s}=24$
GeV, $|y|<3$. Left panel: the GPM results with $\langle q_T^2\rangle=1$ GeV$^2$ are shown by dash-dotted (NRQCD) and dash-double-dotted (ICEM) histograms. Solid and dashed histograms with uncertainty bands are PRA~\cite{Karpishkov:2020wwe} and NLO CPM~\cite{private:1} predictions respectively. Right panel: the GPM predictions in NRQCD (solid histogram with light green uncertainty band) and ICEM (dashed histogram with dark-green uncertainty band) approaches with their uncertainty bands shown.}
\end{figure}

As for $k_{TJ/\psi}$-spectrum, the GPM calculations, both in NRQCD-factorization and ICEM, lead to results consistent
 with NRQCD predictions of Parton Reggeization Approach~\cite{Karpishkov:2020wwe} at small transverse-momentum,
 while the latter predictions are in agreement with NLO NRQCD predictions of Collinear Parton Model~\cite{private:1}
 at high-$k_{TJ/\psi}$ as one can see in the left panel of the Fig.~\ref{fig:NICApt}. Predictions of NRQCD and ICEM
 approaches for $k_{TJ/\psi}$-spectrum are also remarkably consistent with each-other, but ICEM prediction has smaller
 scale-uncertainty (see the right panel of the Fig.~\ref{fig:NICApt}) because the squared matrix element of the hard process
  in ICEM is of $O(\alpha_s^2)$ while for NRQCD approach it is of $O(\alpha_s^3)$. The relative contributions of direct and
   feed-down production at the energy $\sqrt{s}=24$  GeV are given in Tab.~\ref{Tab:DirFD} and they turn out to be consistent with PHENIX data.
    Thus we conclude that we can safely perform predictions for TSSA at NICA energies.

\begin{figure}[h]
\begin{center}
\includegraphics[width=0.4\textwidth, clip=]{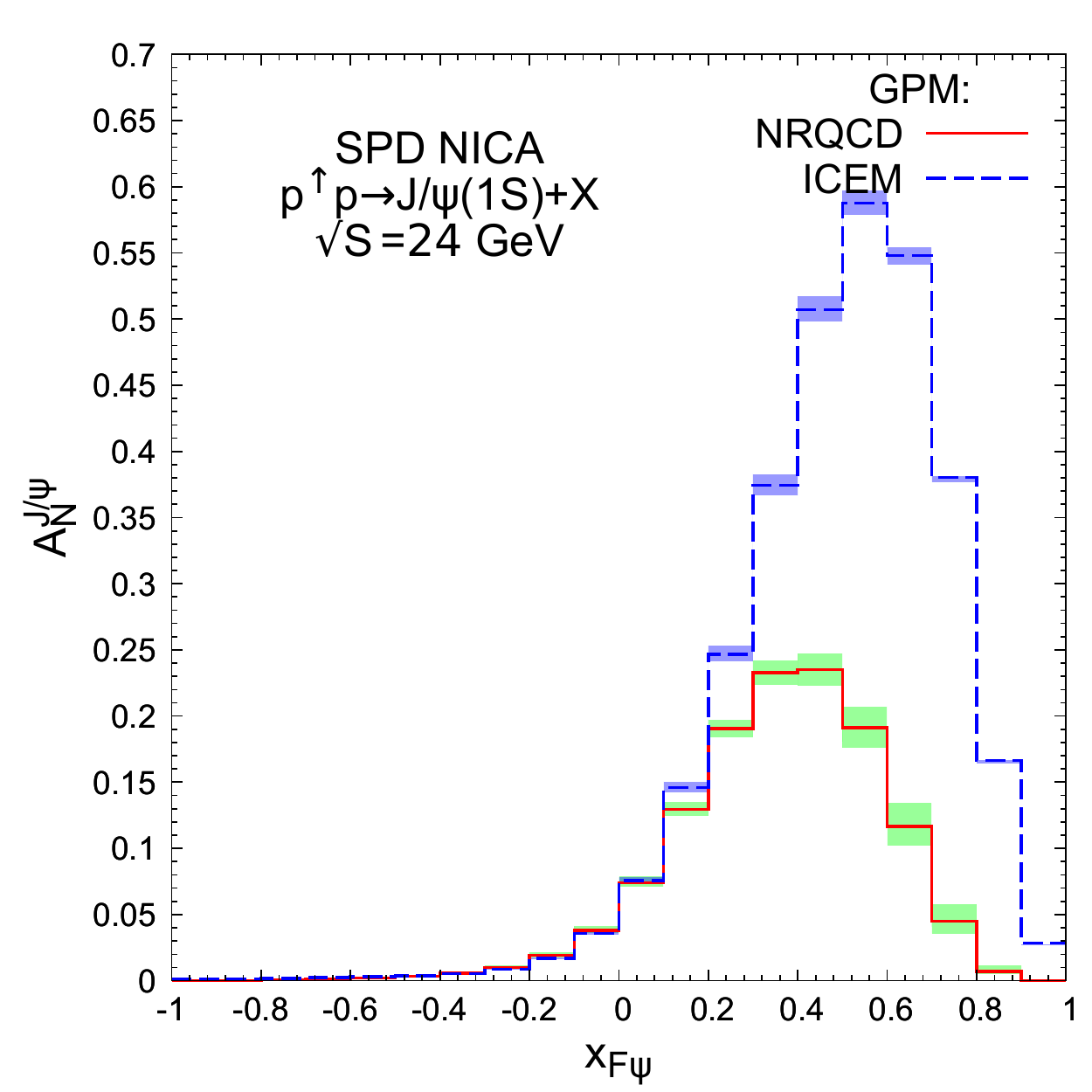}
\includegraphics[width=0.4\textwidth, clip=]{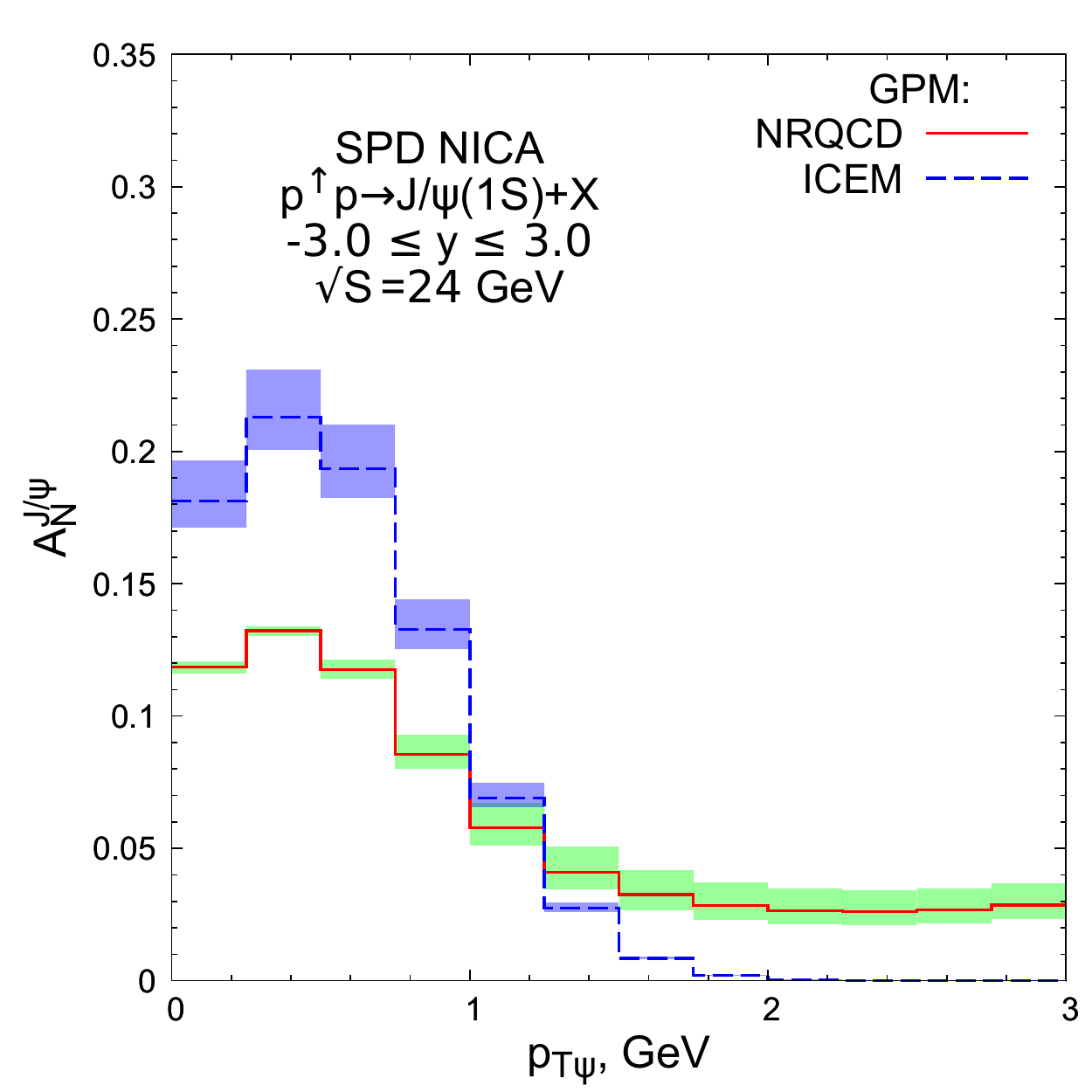}
\end{center}
\caption{\label{fig:AN-NICA-SDIS1} Comparison of predictions in GPM for TSSA $A_N^{J/\psi}$ as function of $x_F$ (left panel) and transverse-momentum (right panel) at $\sqrt{s}=24$ GeV in NRQCD (solid histogram) and ICEM (dashed histogram) approaches. The SDIS1 parametrisation of GSFs is used.}
\end{figure}

\begin{figure}[h]
\begin{center}
\includegraphics[width=0.4\textwidth, clip=]{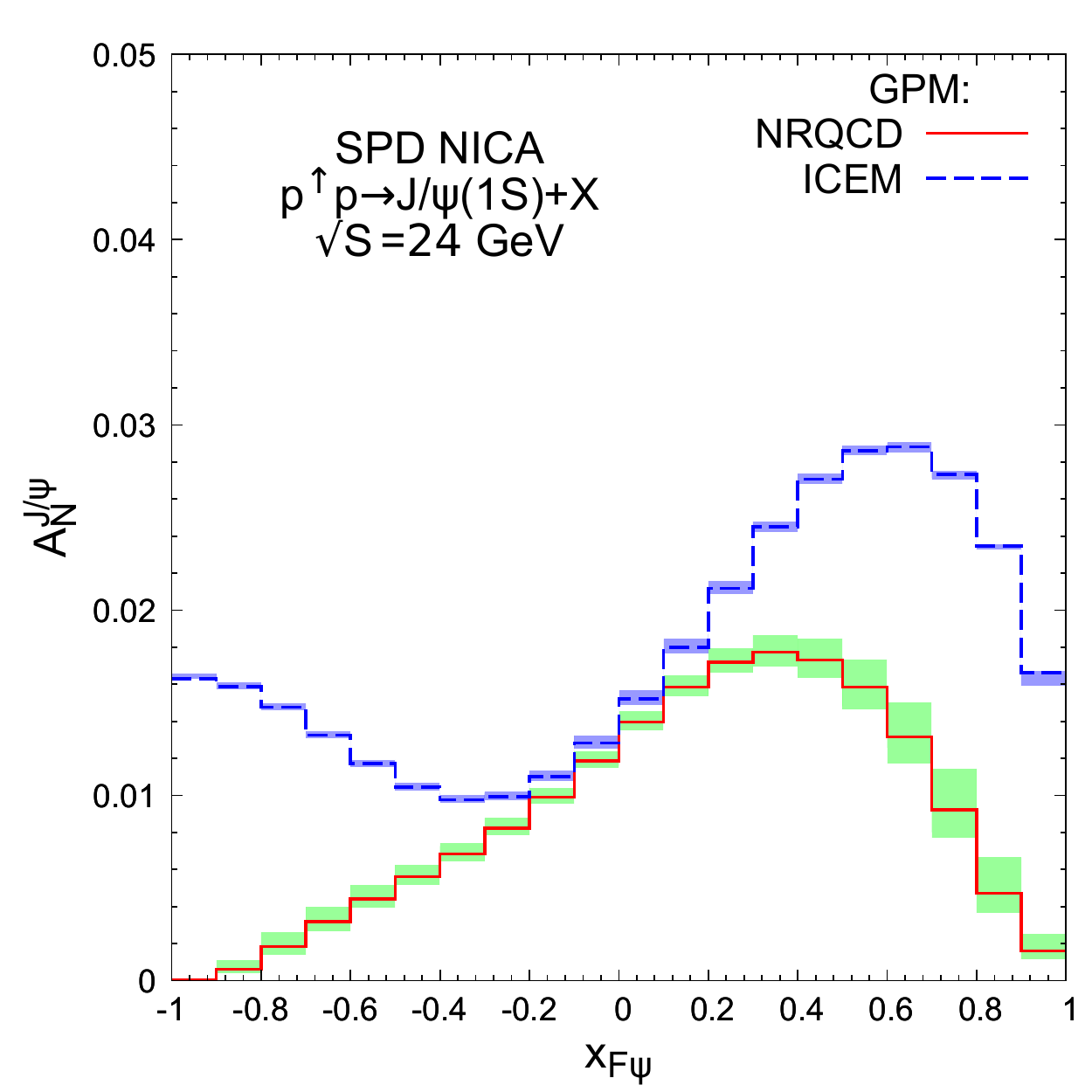}
\includegraphics[width=0.4\textwidth, clip=]{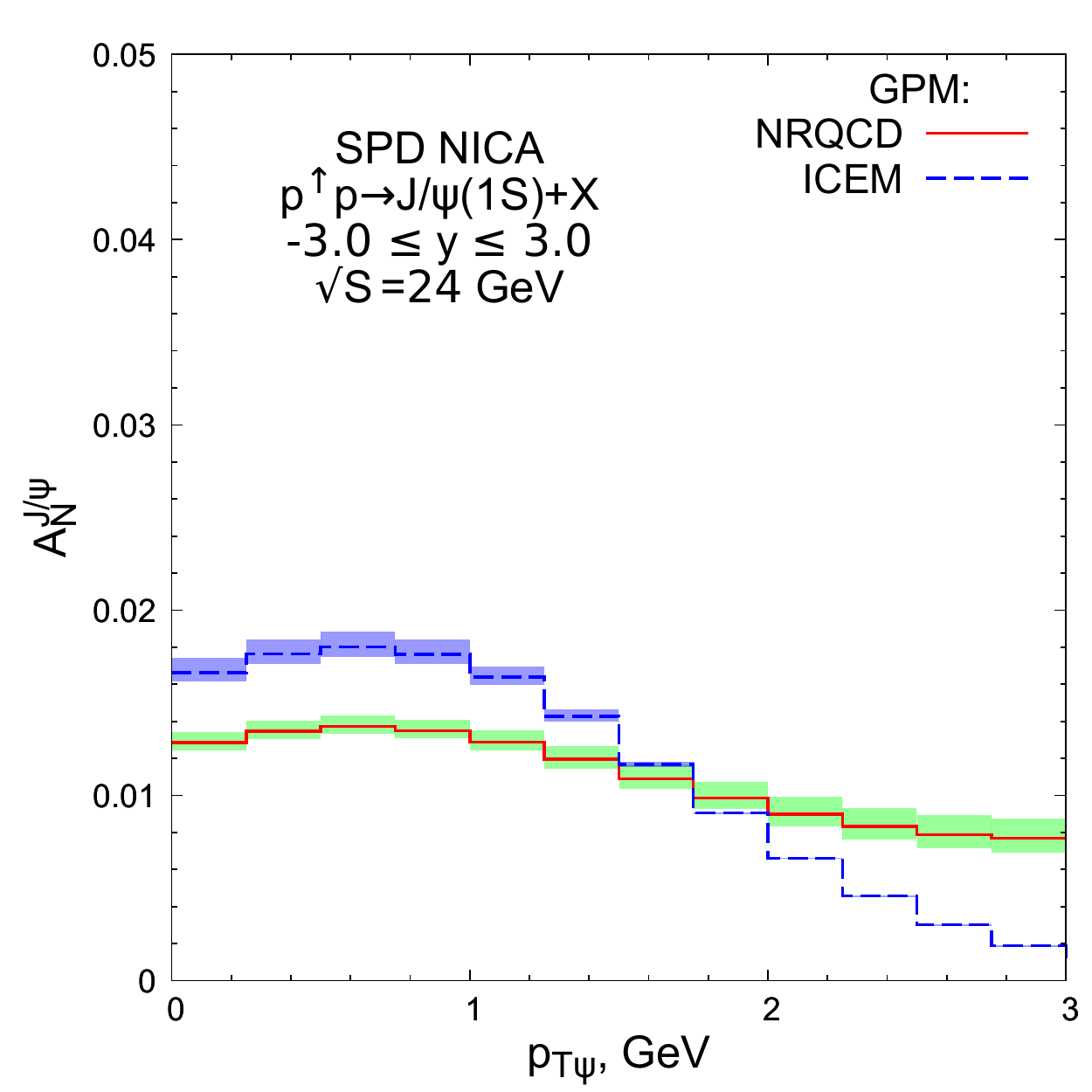}
\end{center}
\caption{\label{fig:AN-NICA-D'Alesio} Comparison of predictions in GPM for TSSA $A_N^{J/\psi}$ as function of $x_F$ (left panel) and transverse-momentum (right panel) at $\sqrt{s}=24$ GeV in NRQCD (solid histogram) and ICEM (dashed histogram) approaches. The D'Alesio {\it et al.} parametrisation of GSFs is used.}
\end{figure}

\begin{figure}[h]
\begin{center}
\includegraphics[width=0.4\textwidth, clip=]{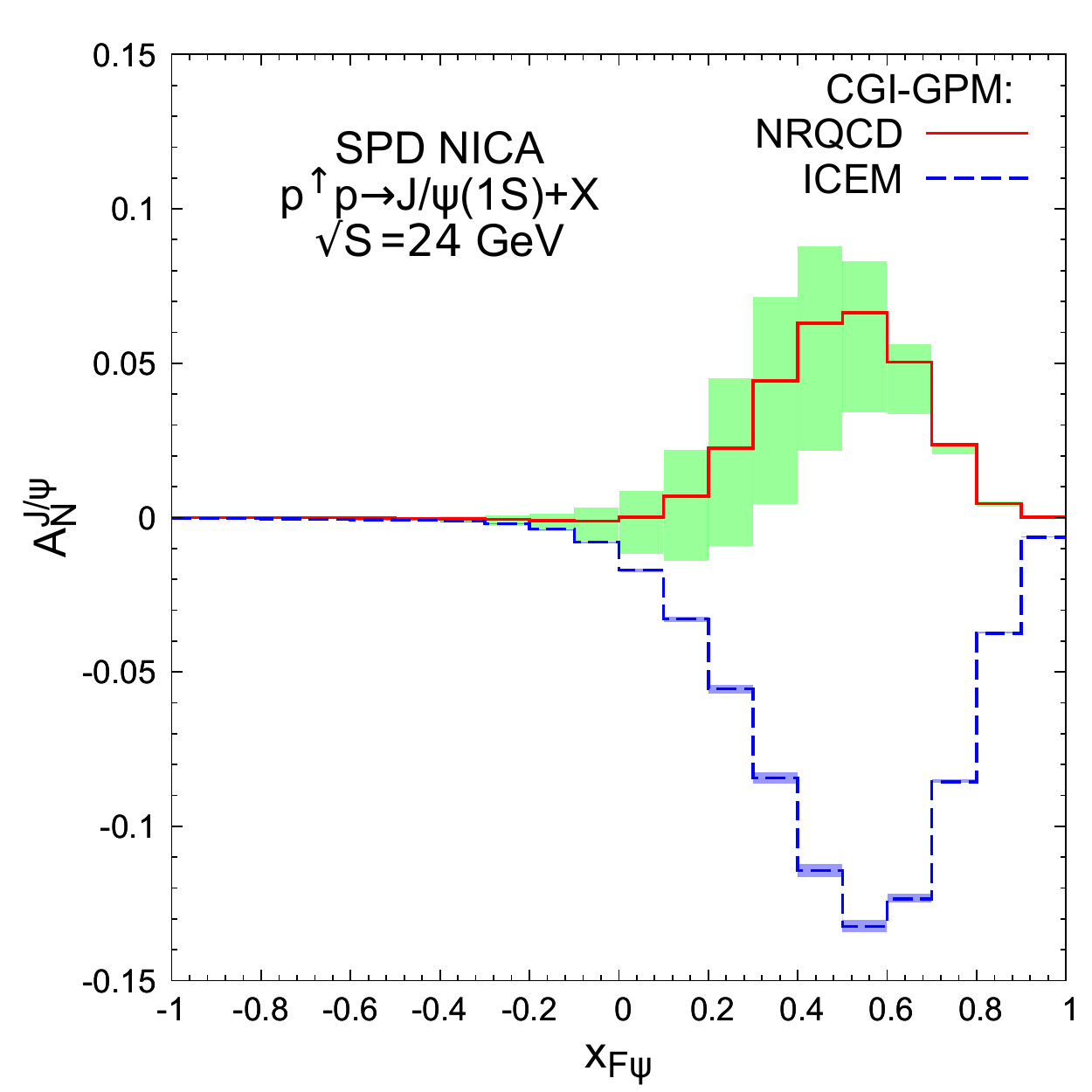}
\includegraphics[width=0.4\textwidth, clip=]{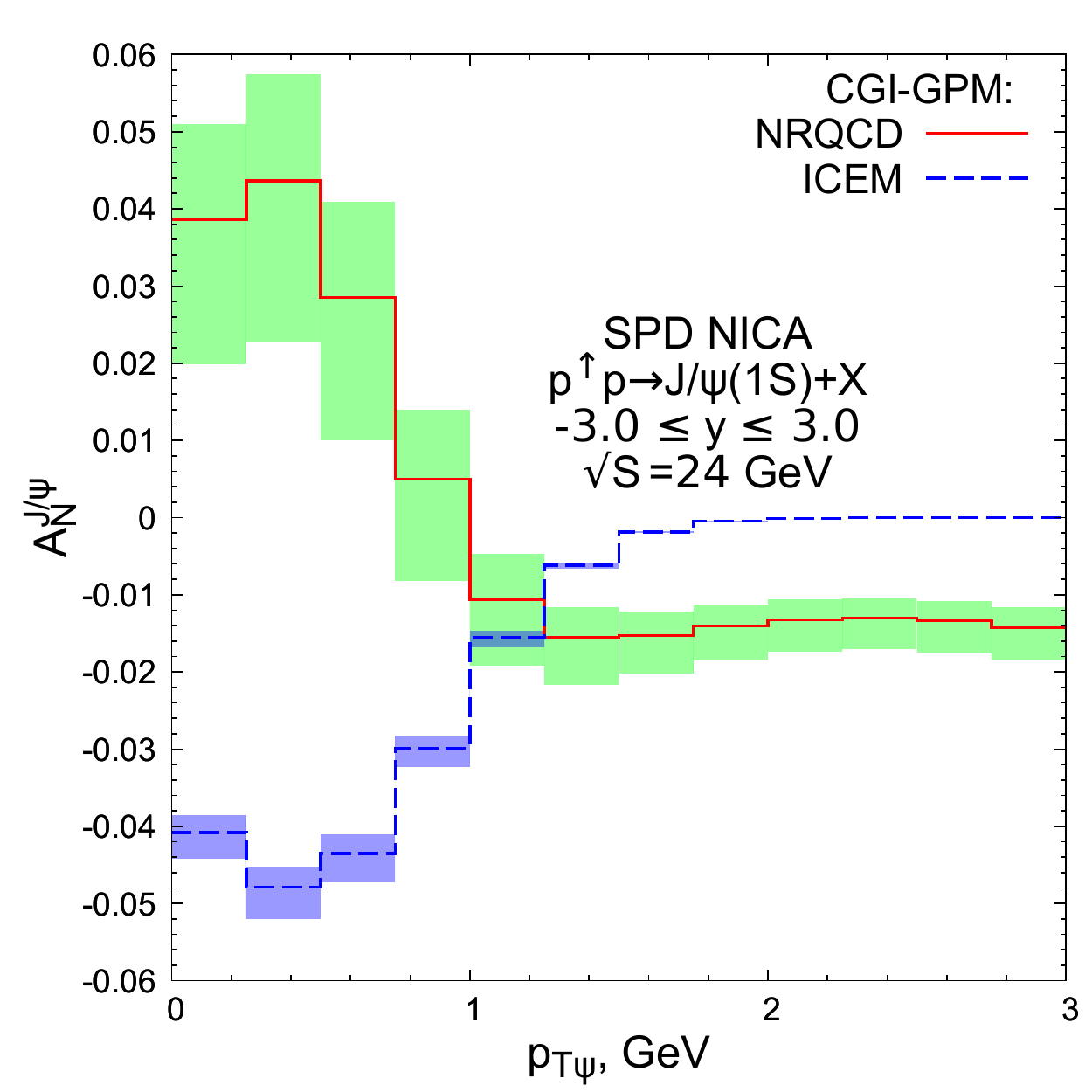}
\end{center}
\caption{\label{fig:CGI-AN-NICA-SDIS1} Comparison of predictions in CGI-GPM for TSSA $A_N^{J/\psi}$ as function of $x_F$ (left panel) and transverse-momentum (right panel) at $\sqrt{s}=24$ GeV in NRQCD (solid histogram) and ICEM (dashed histogram) approaches. The SDIS1 parametrisation of GSFs is used.}
\end{figure}

\begin{figure}[h]
\begin{center}
\includegraphics[width=0.4\textwidth, clip=]{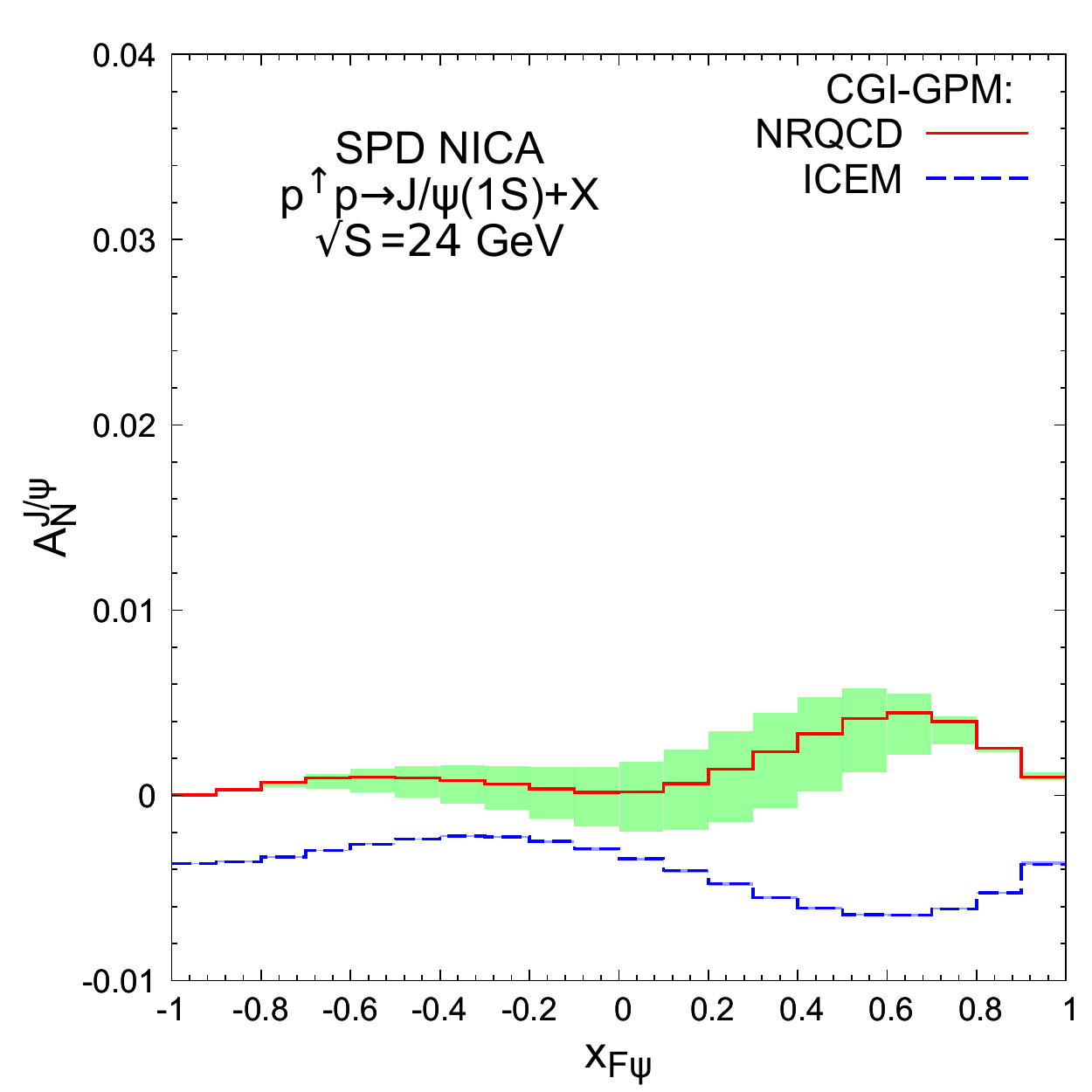}
\includegraphics[width=0.4\textwidth, clip=]{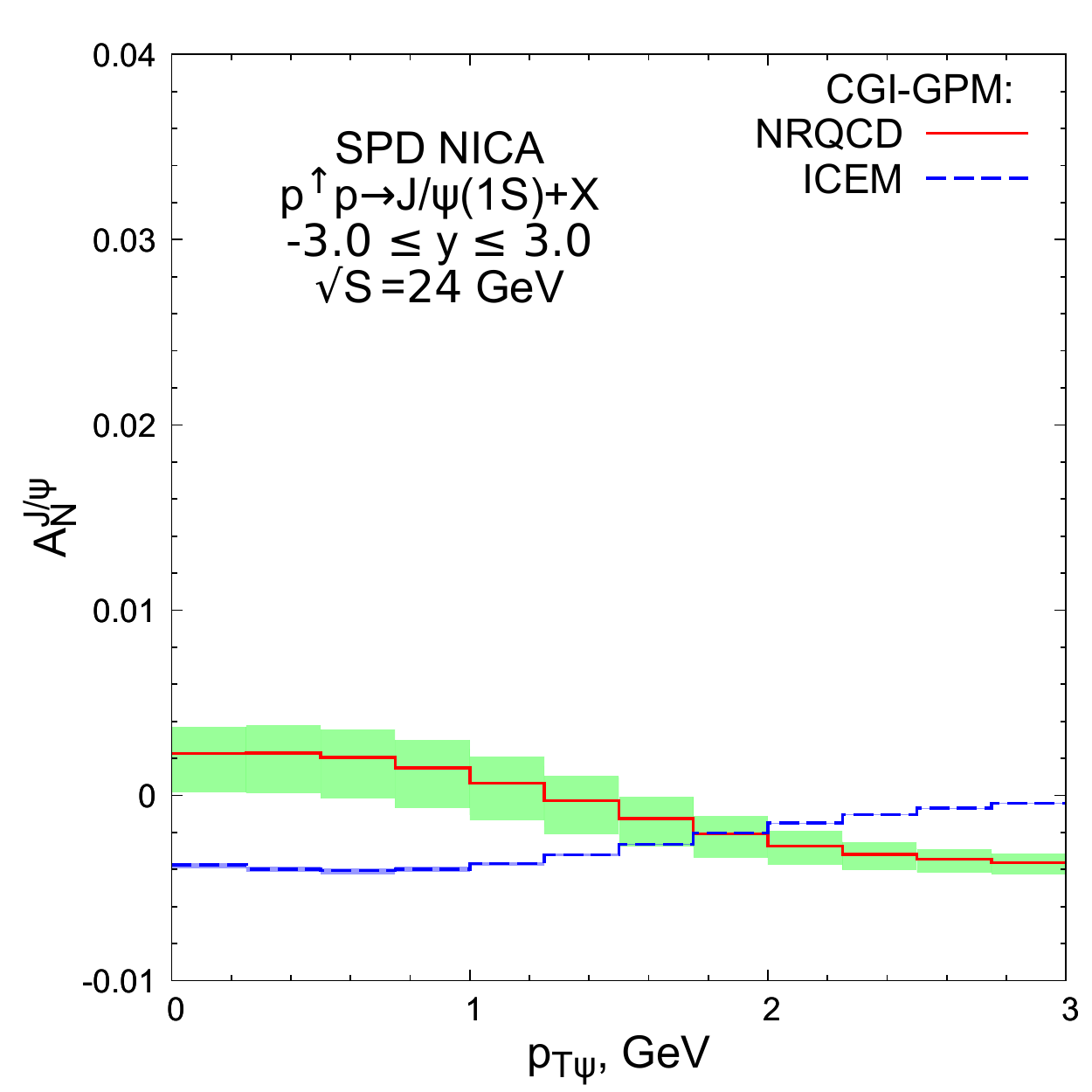}
\end{center}
\caption{\label{fig:CGI-AN-NICA-D'Alesio} Comparison of predictions in CGI-GPM for TSSA $A_N^{J/\psi}$ as function of $x_F$ (left panel) and transverse-momentum (right panel) at $\sqrt{s}=24$ GeV in NRQCD (solid histogram) and ICEM (dashed histogram) approaches. The D'Alesio {\it et al.} parametrisation of GSFs is used.}
\end{figure}

Estimates for TSSA at SPD NICA experiment, computed within NRQCD and ICEM approaches under GPM initial-state factorization assumption are shown in Figs. \ref{fig:AN-NICA-SDIS1} and \ref{fig:AN-NICA-D'Alesio} respectively for the SDIS1 and D'Alesio {\it et al.} parametrisations for GSF. We find that for standard GPM initial-state factorization, the SDIS1 predicts gigantic values for asymmetries at NICA energies -- up to 60\% (Fig.~\ref{fig:AN-NICA-SDIS1}). However such big effects can hardly be expected to appear, since this parametrisation contradicts PHENIX data, when GPM is used. GPM predictions with D'Alesio {\it et al.} parametrisation (Fig.~\ref{fig:AN-NICA-D'Alesio}) look more realistic and they are quite robust against the choice of $J/\psi$-formation model. Measurable asymmetries up to 5\% for the $x_{F\psi}$-spectrum and up to 2\% for $k_{TJ/\psi}$-spectrum are predicted.

 Our results obtained using CGI-GPM initial-state factorization are shown in Figs.~\ref{fig:CGI-AN-NICA-SDIS1} and~\ref{fig:CGI-AN-NICA-D'Alesio} for the SDIS1 and D'Alesio {\it et al.} parametrisations, correspondingly. As it was for the case of PHENIX kinematics discussed above, the smaller in absolute value TSSAs of charmonium production are predicted within CGI-GPM factorization in comparison to the usual GPM factorization. Also in the Figs.~\ref{fig:CGI-AN-NICA-SDIS1} and~\ref{fig:CGI-AN-NICA-D'Alesio}, within the CGI-GPM+CSM model we observe sign-change of the $A_N$ for $p_T\approx 1$~GeV, similar to observations in the Ref.~\cite{DAlesio:2020eqo}. This sign-change happens mostly due to a negative color factor in Eq.~(\ref{eq:ISIcolfact}) and a large contribution of direct $J/\psi$ production (see Tab.~\ref{Tab:DirFD}). Another interesting observation is, that the ICEM predicts only negative values for the TSSA within CGI-GPM, because integrated coefficient function~(\ref{eq:totCSCGI}) is negative for $0\leq w \leq 1$. Finally, from Figs.~\ref{fig:CGI-AN-NICA-SDIS1} and~\ref{fig:CGI-AN-NICA-D'Alesio} one can see that the CSM and the ICEM predict values of the TSSA opposite in sign for SPD NICA kinematic conditions. This potentially allows to discriminate between these two approaches of hadronisation within the CGI-GPM, if the energy scan from $\sqrt{s}=10$ to $27$ GeV will be performed, allowing to disentangle between effects of initial and final-state factorization.

\section*{Conclusions}
 \label{sec:Con}

In the present paper we have performed a phenomenological analysis of gluon Sivers function contribution to the transverse TSSA of prompt $J/\psi$-production within NRQCD-factorization (essentially Color-Singlet Model in our case) and ICEM for the description of $J/\psi$-formation, employing both state-of-art initial-state factorization models: GPM and CGI-GPM. The goal of our analysis was to make predictions for TSSA in the kinematic conditions of planned SPD NICA experiment. We have found, that within standard GPM initial-state factorization, the SDIS1 parametrisation for gluon Sivers function contradicts PHENIX data, while parametrisation of D'Alesio {\it et al.} leads to reasonable predictions for magnitude, $J/\psi$ transverse-momentum and $x_{F\psi}$ dependence of the asymmetry with $|A_N|\lesssim 2-3$\% (Fig.~\ref{fig:AN-NICA-D'Alesio}). Within CGI-GPM initial-state factorization, contradiction of SDIS1-parametrization with PHENIX data is eliminated, and it predicts $|A_N|\lesssim 5-10$\% at SPD NICA kinematics (Fig.~\ref{fig:CGI-AN-NICA-SDIS1}). Hence, observation of sizable transverse TSSA in inclusive $J/\psi$-production does not contradict existing experimental data and their theoretical interpretation within a wide range of $J/\psi$-formation and initial-state factorization models. In any case, measurements at SPD NICA will significantly constrain our knowledge about gluon Sivers function in a proton.

\section*{Acknowledgments}
 Authors are grateful to Mathias Butensch\"on and Bernd Kniehl for providing their NLO CPM predictions
  for $J/\psi$ transverse-momentum spectrum at NICA, as well as to Igor Denisenko, Alexey Guskov, Oleg Teryaev
   and other members of SPD NICA Collaboration for useful and encouraging physics discussions.
   The work has been supported in parts by the Ministry of science and higher education of Russia via State assignment to
   educational and research institutions under
project FSSS-2020-0014 and by the Foundation for the Advancement of
Theoretical Physics and Mathematics BASIS, grant No. 18-1-1-30-1.



\end{document}